\documentclass[12pt]{article}

\usepackage{amsmath}
\usepackage{amssymb}

\usepackage{graphicx}
\usepackage{float}
\usepackage{ulem}
\usepackage{comment}

\def \E{\mathbb{E}}

\def \R{\mathbb{R}}

\def \Lc{{\cal L}}

\def \eps{\varepsilon}

\def\reff#1{{\rm(\ref{#1})}}

\def \ind{1\!\!1}

\def \benu {\begin{enumerate}}
\def \eenu {\end{enumerate}}

\def\beqs{\begin{eqnarray*}}
\def\enqs{\end{eqnarray*}}
\def\beq{\begin{eqnarray}}
\def\enq{\end{eqnarray}}

\newtheorem{lem}{Lemma}
\newtheorem{theo}{Theorem}
\newtheorem{rem}{Remark}
\newtheorem{pro}{Proposition}
\newtheorem{coro}{Corollary}
\newtheorem{hyp}{Assumption}

\newenvironment{pf}{\ \\ {\bf Proof: }}{\hfill\mbox{$\diamond$}\medskip}

\begin{document}
\vspace{1.5cm}

\author{Erwan Pierre\thanks{EDF R\&D OSIRIS. Email:  \sf erwan.pierre@edf.com}~~~ St\'{e}phane
Villeneuve\thanks{Toulouse School of Economics (CRM-IDEI),
Manufacture des Tabacs, 21, All\'ee de Brienne, 31000 Toulouse,
France.
 Email: {\sf
stephane.villeneuve@tse-fr.eu}. This authors gratefully acknowledges the financial support of the research initiative IDEI-SCOR "Risk Market and Creation Value" under the aegis of the risk foundation. }~~~Xavier Warin\thanks{\small{EDF R\&D \& FiME}, Laboratoire de Finance des
March\'es de l'Energie (www.fime-lab.org)
}
}

\title{Liquidity Management with Decreasing-returns-to-scale and Secured Credit Line.}
 \vspace{1cm}\renewcommand{\today}

\maketitle

\noindent{\bf Abstract:} 
This paper examines the dividend and investment policies of a cash constrained firm, assuming a decreasing-returns-to-scale technology and adjustment costs. We extend the literature by allowing the firm to draw on a secured credit line both to hedge against cash-flow shortfalls and to invest/disinvest in productive assets.
We formulate this problem as a bi-dimensional singular control problem and use both a viscosity solution approach and a verification technique to get qualitative properties of the value function. We further solve quasi-explicitly the control problem in two special cases.\\

\noindent{\it Keywords: Investment, dividend policy, singular control, viscosity solution} \\
\noindent{\it JEL Classification numbers:} C61; G35.\\
\noindent{\it MSC Classification numbers:} 60G40; 91G50; 91G80.

\section{Introduction}
In a world of perfect capital market, firms could finance their operating costs and investments by issuing shares at no cost. As long as the net present value of a project is positive, it will find investors ready to supply funds. This is the central assumption of the Modigliani and Miller theorem \cite{mm:58}. On the other hand, when firms face external financing costs, these costs generate a precautionary demand for holding liquid assets and retaining earnings. This departure from the Modigliani-Miller framework has received a lot of attention in recent years and has given birth to a serie of papers explaining why firms hold liquid assets. Pioneering papers  are Jeanblanc and Shiryaev \cite{js:95}, Radner and Shepp \cite{rs:96} while more recent studies include Bolton, Chen and Wang \cite{bcw:11},  D\'ecamps, Mariotti, Rochet and Villeneuve \cite{dmrv:11} and  Hugonnier, Malamud and Morellec \cite{hmm:11}. In all of these papers, it is assumed that firms are all equity financed. Should it runs out of liquidity, the firm either liquidates or raises new funds in order to continue operations by issuing equity. This binary decision only depends on the severity of issuance costs.\\
The primary objective of our paper is to study a setup where a cash-constrained firm has a mixed capital structure. To do this, we build on the paper by Bolton, Chen and Wang \cite{bcw:11} chapter V to allow the firm to access a secured credit line. While \cite{bcw:11} assumed a constant-returns-to-scale and  homogeneous adjustment costs which allows them to work with the firm's cash-capital ratio and thus to reduce the dimension of their problem, we rather consider a decreasing-returns-to-scale technology with linear adjustments costs.\\
Bank credit lines are a major source of liquidity provision in much the same way as holding cash does. 
Kashyap, Rajan and Stein \cite{KRS:02} found that 70\% of bank borrowing by US small firms is through credit line. However, access to credit line is contingent to the solvency of the borrower which makes the use on credit line costly through the interest rate and thus makes it an imperfect substitute for cash (Sufi \cite{S:09}). From a theoretical viewpoint, the use of credit lines can be justified by moral hazard problems (Holmstrom-Tirole \cite{ht:98}) or from the fact that banks can commit to provide liquidity to firms when capital market cannot because banks have better screening and monitoring skills (Diamond \cite{D:84})\\
In this paper, we model credit line as a full commitment lending relationship between a firm and a bank. The lending contract specifies that the firm can draw on a line of credit as long as its outstanding debt, measured as the size of the firm's line of credit, is below the value of total assets (credit limit).  The liability side of the balance sheet of the firm consists in two different types of owners: shareholders and bankers. Should the firm be liquidated, bankers have seniority over shareholders on the total assets. 
We assume that the secured line of credit continuously charges a variable spread\footnote{The spread may be justified by the cost of equity capital for the bank. Indeed, the full commitment to supply liquidity up to the firm's credit limit prevents bank's shareholders to allocate part of their equity capital to more valuable investment opportunities. } over the risk-free rate $r$ indexed on the firm's outstanding debt, the higher the size of firm's line of credit, the higher the spread is. With this assumption, the secured line of credit is somehow similar to the performance-sensitive debt studied in \cite{mst:10} except that the shareholders are here forced to go bankrupt when they are no more able to secure the credit line with their assets.

 Many models initiated by Black and Cox \cite{bc:76} and Leland \cite{le:94} that consider the traditional tradeoff between tax and bankruptcy costs as an explanation for debt issuance study firms liabilities as contingent claims on its underlying assets, and bankruptcy as an endogenous decision of the firm management. On the other hand, these models assume costless equity issuance and thus put aside liquidity problems. As a consequence, the firm's decision to borrow on the credit market is independent from liquidity needs and investment decisions. A notable exception is a recent paper by Della Seta, Morellec, Zucchi \cite{DSMZ:15} which studies the effects of debt structure and liquid reserves on banks' insolvency risk. Our model belongs to the class of models that consider endogenous bankruptcy of a firm with mixed capital structure replacing taxes with liquidity constraints.\\
From a mathematical point of view, problems of cash management have been formulated as singular stochastic optimal control problems. As references for the theory of singular stochastic
control, we may mention the pioneering works of Haussman and Suo \cite{hs:95a} and \cite{hs:95b} and for application to cash management problems H\o jgaard and Taksar \cite{ht:99},
Asmussen, H\o jgaard and Taksar \cite{aht:00}, Choulli, Taksar and
Zhou \cite{ctz:03}, Paulsen \cite{pa:03} among others. To merge corporate liquidity, investment and financing in a tractable model is challenging because it involves a rather difficult three-dimensional singular control problem with stopping where the  state variables are the book value of equity, the size of productive asset and the size of the firm credit line while the stopping time is the decision to default. The literature on multi-dimensional control problems relies mainly on the study of leading examples. A seminal example is the so-called finite-fuel problem introduced by Benes, Shepp and Witsenhausen
\cite{bsw:80}. This paper provides a rare example of a bi-dimensional optimization problem that combines singular
control and stopping that can be solved explicitly by analytical means. More recently, Federico and Pham \cite{fp:14} have solved a degenerate bi-dimensional singular control problem to study a reversible investment problem where a social planner aims to control its capacity production in order to fit optimally the random demand of a good. Our paper complements the paper by Federico and Pham \cite{fp:14} by introducing firms that are cash-constrained\footnote{Ly Vath, Pham and Villeneuve \cite{lpv:08} have also studied a reversible investment problem in two alternative technologies for a cash-constrained firm that has no access to external funding}. To our knowledge, this is the first time that such a combined approach is used. This makes the problem much more complicated and we do not pretend solving it with full generality,  but rather, we pave the way for future developments of these multidimensional singular control models. In particular, we lose the global convexity property of the value function that leads to the necessary smooth-fit property in \cite{fp:14} (see Lemma \ref{lem7}). Instead, we will give properties of the value function (see Proposition \ref{prop_cont}) and characterize it by means of viscosity solution (see Theorem \ref{charac_viscosity}). Furthermore, we will solve explicitly by a standard verification argument the peculiar case of costless reversible investment. A last new result is our characterization of the endogenous bankruptcy in terms of the profitability of the firm and the spread function.

The remainder of the paper is organized as follows. Section 2 introduces the model with a productive asset of fixed size, formalizes the notion of secured line of credit and defines the shareholders value function. Section 3 contains our first main result, it describes the optimal credit line policy and gives the analytical characterization of the value function in terms of a free boundary problem for a fixed size of productive assets. Section 4 is a technical section that builds the value function by solving explicitly the free boundary problem. Section 5 extends the analysis to the case of reversible investment on productive assets and paves the way to a complete characterization of the dividend and investment policies.

\section{The No-investment Model}

We consider a firm owned by risk-neutral shareholders, with a productive asset of fixed size $K$, whose price is normalized to unity, that has an agreement with a bank for a secured line of credit. 
The credit line is a source of funds available at any time up to a credit limit defined as the total value of assets. The firm has been able to secure the credit line by posting its productive assets as collateral. Nevertheless, in order to make the credit line attractive for bank's shareholders that have dedicated part of their equity to this agreement, we will assume that the firm will pay a variable spread over the risk-free rate $r$ depending on the size of the used part of the credit line. In this paper, the credit line contract is given and thus the spread is exogenous, see Assumption \ref{hyp1}. Finally, building on Diamond's result \cite{D:84} we assume that the costs of equity issuance are so high that the firm is unwilling to increase its cash reserves by raising funds in the equity capital market and prefers drawing on the credit line. The firm is characterized at each date $t$ by the following balance sheet:

\begin{center}
{\normalsize
\begin{tabular}{|c|c|}
\hline
&  \\
$\;\; K \;\;$ & $\;\; X_t \;\;$ \\
&  \\
$\;\; M_t \;\;$ & $\;\; L_t \;\;$ \\
&  \\ \hline
\end{tabular}
}
\end{center}

\begin{itemize}
\item $K$ represents the firm's productive assets, assumed to be constant\footnote{The extension to the case of variable size will be studied in Section 4} and normalized to one.

\item $M_t$ represents the amount of cash reserves or liquid assets.

\item $L_t$ represents the size
of the credit line, i.e. the amount of cash that has been drawn on the line of credit.
 
\item Finally, $X_t$ represents the book value of equity.
\end{itemize}

The productive asset continuously generates cash-flows over time. The cumulative cash-flows process $R=(R_t)_{t\ge 0}$ is modeled as an arithmetic Brownian motion with drift $\mu$ and volatility $\sigma$ which is defined over a complete probability space $(\Omega, \mathcal{F}, \mathbb{P})$ equipped with a filtration $(\mathcal{F}_t)_{t \geq 0}$. Specifically, the cumulative cash-flows evolve as
$$
dR_t=\mu\,dt+\sigma\,dB_t
$$
where $(B_t)_{t \geq 0}$ is a standard one-dimensionnal Brownian motion with respect to the filtration $(\mathcal{F}_t)_{t \geq 0}$.\\
 
Credit line requires the firm to make an interest payment that is increasing in the size of the used part of the credit line. We assume that the interest payment is defined by a function $\alpha(.)$ where
\begin{hyp}\label{hyp1}
$\alpha$ is a strictly increasing, continuously differentiable  convex function such that
\begin{equation}\label{controlederiveealpha}
\forall x \ge 0, \alpha'(x) \geq r
\hbox{ and }
\alpha(0) = 0.
\end{equation}
The credit line spread $\alpha(.)-r$ is thus strictly positive and increasing.
\end{hyp}

The liquid assets earn a rate of interest $r-\delta$ where $\delta \in (0,r]$ represents a carry cost of liquidity\footnote{This assumption is standard in models with cash. It captures in a simple way the agency costs, see \cite{dmrv:11}, \cite{hmm:11} for more details}. Thus, in this framework, the cash reserves evolve as
\begin{equation}
dM_{t} = (r-\delta)M_{t^-}dt+(\mu-\alpha(L_{t^-})) dt + \sigma dB_{t} - dZ_{t} + dL_t
\end{equation}
where $(Z_t)_t$ is an increasing right-continuous $({\cal F}_t)_t$ adapted process representing the cumulative dividend payment up to time $t$ and $(L_t)_t$ is a positive right-continuous $({\cal F}_t)_t$ adapted process representing the size of the credit line (outstanding debt) at time $t$. Using the accounting relation $1+M_t=X_t+L_t$, we deduce the dynamics for the book value of equity
\begin{equation}\label{bookvalueequitydynamics}
dX_{t} = (r-\delta)X_{t-}dt+(\mu-(r-\delta)+(r-\delta)L_{t-}-\alpha(L_{t^-})) dt + \sigma dB_{t} - dZ_{t}.
\end{equation}
Finally, we assume the firm is cash-constrained in the following sense:
\begin{hyp} \label{hyp2}
The cash reserves must be non negative and the firm management is forced to liquidate when the book value of equity hits zero. Using the accounting relation, this is equivalent to assume bankers get back all the productive assets after bankruptcy.
\end{hyp}
The goal of the management is to maximize shareholders value which is defined as the expected discounted value of all future dividend payouts. Because shareholders are assumed to be risk-neutral, future cash-flows  are discounted at the risk-free rate $r$. The firm can stop its activity at any time by distributing all of its assets to stakeholders. Thus, the objective is to maximize over the admissible control $\pi = (L,Z)$ the functional
\[
V(x,l; \pi) =\mathbb{E}_{x,l}\left( \int_{0}^{\tau_0}{e^{-r t}dZ_{t}}\right)
\]
where 
$$
\tau_0=\inf\{t \ge 0, X_t^{\pi} \le 0\}
$$
according to Assumption \ref{hyp2}. Here $x$ (resp. $l$) is the initial value of equity capital (resp. debt). 
We denote by $\Pi$ the set of admissible control variables and define the shareholders value function  by 
\begin{equation}\label{value}
V^{*}(x,l) = \sup_{\pi \in \Pi} V(x,l;\pi).
\end{equation}

\begin{rem}
We suppose that the cash reserves must be non negative (Assumption \ref{hyp2}) so to be admissible, a control $\pi = (L, Z)$ must satisfy at any time $t$
\[
dZ_t \leq X_{t^-}.
\]

\end{rem}
\section{No-investment Model solution}
This section derives the shareholders value and the optimal dividend and credit line policies. It relies on a standard HJB characterization of the control problem and a verification procedure.
\subsection{Optimal credit line issuance}
The shareholders’ optimization problem  (\ref{value}) involves two state variables, the value of equity capital $X_t$ and the size of the credit line $L_t$, making its resolution difficult. Fortunately, the next proposition will enable us to reduce the dimension and make it tractable the computation of $V^*$. Proposition \ref{reduction} shows that credit line issuance is only optimal when the cash reserves are depleted.

\begin{pro} \label{reduction}
A necessary and sufficient condition to draw on the credit line is that the cash reserves are depleted, that is 
\[
\forall t \in \mathbb{R}^{+}, L_tM_t=0 \hbox{ or equivalently }L_t=(1-X_t)^+.
\] 
\end{pro}

\begin{pf}
First, by Assumption \ref{hyp2}, it is clear that the firm management must draw on the credit line when cash reserves are nonpositive.  
Conversely, assume that the level of cash reserves $m$ is strictly positive. We will show that it is always better off to reduce the level of outstanding debt by using the cash reserves. We will assume that the initial size of the credit line is $L_{0-}=l > 0$ and denote $\pi_t=(L_t,Z_t)$ any admissible strategy. Let us define by $\phi$ the cost of the credit line on the variation of the book value of equity, that is $\phi(l)=\alpha(l)-(r-\delta)l$ such that the book value of equity dynamics is
\begin{equation}\label{bookvalueequitydynamics2}
dX_{t} = (r-\delta)X_{t-}dt+(\mu-(r-\delta)-\phi(L_{t^-})) dt + \sigma dB_{t} - dZ_{t}.
\end{equation}
Note that $\phi$ is strictly increasing. We first assume that the firm does not draw on the credit line at time 0, $L_0=l$. Because $m>0$, we will built a strategy from $\pi$ as follows:
$$
\left\{
\begin{array}{ll}
L_0^\epsilon &=l -\epsilon \hbox{ for } 0<\epsilon <\min(m,l) \hbox{ and } 0 \le L_t^\epsilon \le L_t,\\
Z_t^\epsilon &=Z_t+\int_0^t\left( \phi(L_s) -\phi(L_s^\epsilon)\right)\,ds
\end{array}
\right.
$$
Note that the credit line issuance strategy $L^\epsilon$ consists in always having less debt that under the credit line issuance strategy $L$ and because $\phi$ is increasing, the dividend strategy $Z_t^\epsilon$ pays more than the dividend strategy $Z_t$. Furthermore, denoting by $\pi^\epsilon=(L_t^\epsilon,Z_t^\epsilon)$, equation \eqref{bookvalueequitydynamics2} shows that the bankruptcy time under $\pi^\epsilon$ starting from $(x,l-\epsilon)$ and the bankruptcy time under $\pi$ starting from $(x,l)$  have the same distribution. Therefore, 

\begin{eqnarray*}
V(x,l;\pi^\epsilon)&=& \E_{(x,l-\epsilon)}\left(\int_0^{\tau_0^{\pi^\epsilon}} e^{-rs}\,dZ_s^\epsilon\right)\\
& > &\E_{(x,l-\epsilon)}\left(\int_0^{\tau_0^{\pi^\epsilon}} e^{-rs}\,dZ_s\right)\\
&=& \E_{(x,l)}\left(\int_0^{\tau_0^{\pi}} e^{-rs}\,dZ_s\right)\\
&=& V(x,l;\pi),
\end{eqnarray*}
which shows that it is better off to follow $\pi^\epsilon$ than $\pi$.
So if $m>l$, it is optimal to set $l=0$ by using $m-l$ units of cash reserves while if $m<l$, it is optimal to reduce the debt to $l-m$. In any case, at any time $L_t = (1-X_t)^+$.\\
Now, if we assume that the firm draw on the credit line at time 0, i.e.  $\Delta L_0 \neq 0$, two cases have to be considered.
\begin{itemize}
\item $L_0=0$ which is possible only if $m>l$. In that case, we set $L^\eps_t=L_t$ and $Z^\eps_t=Z_t$ for $t>0$.
\item $L_0>0$. In that case, we take the same strategy $\pi^\epsilon$ with $0<\epsilon <\min(m,l+\Delta L_0)$. 
\end{itemize}
\end{pf}\\
According to Proposition \ref{reduction}, we define the value function as $
v^{*}(x) = V^*(x,(1-x)^+)$. The rest of the section is concerned with the derivation of $v^*$.  

\subsection{Analytical Characterization of the firm value}

Because the level of capital is assumed to be constant, Proposition \ref{reduction} makes our control problem one-dimensional. Thus, we will follow a standard verification procedure to characterize the value function in terms of a free boundary problem. In order to focus on the impact of credit line on the liquidity management, we will assume hereafter that $\delta=r$. This assumption is without loss of generality but allow us to be more explicit in the analytical derivation of the HJB free boundary problem. We denote by ${\cal L}$ the differential operator:
\begin{equation}
{\cal L} \Phi = (\mu - \alpha((1-x)^+))\Phi^{'}(x) + \frac{\sigma^2}{2}\Phi^{''}(x)-r \Phi.
\end{equation}
We start by providing the following standard result which establishes that a smooth solution to a free boundary problem coincides with the value function $v^*$. 
\begin{pro} \label{propositionFB}
Assume there exists a $C^1$ and piecewise twice differentiable function $w$ on $(0,+\infty)$ together with a pair of constants $(a,b) \in \mathbb{R}^+\times\mathbb{R}^+$ such that,
\begin{equation}
\begin{split}
\forall x \leq a,& \qquad  {\cal L} w \leq 0 \text{ and } w(x) = x\\
\forall a \leq x \leq b,& \qquad {\cal L} w = 0 \text{ and } w'(x) \geq 1\\
\forall x > b,& \qquad {\cal L} w \leq 0 \text{ and } w'(x) = 1.\\
\end{split}
\end{equation}

\begin{equation} \label{smoothfit}
\hbox{ with } w^{''}(b) = 0
\end{equation}
then $w=v^*$.
\end{pro}

\begin{pf}
Fix a policy $\pi=(Z) \in \Pi$. Let :
\[
dX_{t} = (\mu-\alpha((1-X_{t^-})^+) dt + \sigma dB_{t} - dZ_{t}, \qquad X(0^-) = x
\]
be the dynamic of the book value of equity under the policy $\pi$. Let us decompose $Z_t=Z_t^c+\Delta Z_t$ for all $t \ge 0$ where $Z_t^c$ is the continuous part of $Z$.\\
Let $\tau_\eps$ the first time when $X_t=\eps$. Using the generalized It\^o's formula, we have :
\[
\begin{split}
e^{-r(t\wedge \tau_\eps)} w(X_{t \wedge \tau_\eps}) = w(x) &+ \int_0^{t \wedge \tau_\eps}e^{-r s}{\cal L}w(X_s)ds + \int_0^{t \wedge  \tau_\eps}\sigma e^{-r s}w^{'}(X_s)dB_s\\
&-\int_0^{t \wedge \tau_\eps}e^{-r s}w^{'}(X_s)dZ^c_s\\
&+\sum_{0 \leq s \leq t \wedge \tau_\eps}e^{-r s}[w(X_s) - w(X_{s^-})].\\
\end{split}
\]
Because $w'$ is bounded, the third term is a square integrable martingale. Taking expectation, we obtain
\[
\begin{split}
w(x) = \textbf{}& \mathbb{E}_x[e^{-r(t\wedge \tau_\eps)} w(X_{t \wedge \tau_\eps})] - \mathbb{E}_x\left[\int_0^{t \wedge \tau_\eps}e^{-r s}{\cal L}w(X_s)ds\right]\\
&+ \mathbb{E}_x\left[ \int_0^{t \wedge \tau_\eps}e^{-r s}w^{'}(X_s)dZ_s^c\right]\\
&- \mathbb{E}_x\left[\sum_{0 \leq s \leq t \wedge \tau_\eps}e^{-r s}[w(X_s) - w(X_{s^-})]\right].\\
\end{split}
\]
Because $w' \geq 1$, we have $w(X_s) - w(X_{s^-}) \leq \Delta X_s = -\Delta Z_s$ therefore the third and the fourth terms are bounded below by
$$
\mathbb{E}_x\left(\int_0^{t \wedge \tau_\eps} e^{-r s}w^{'}(X_s)dZ_s\right).
$$
Furthermore $w$ is positive because $w$ is increasing with $w(0) = 0$ and ${\cal L}w \leq 0$ thus the first two terms are positive. Finally, 
\[
w(x) \ge \mathbb{E}_x\left(\int_0^{t \wedge \tau_\eps} e^{-r s}w^{'}(X_s)dZ_s\right) \geq \mathbb{E}_x\left(\int_0^{t \wedge  \tau_\eps} e^{-r s}dZ_s\right).
\]
Letting t $\rightarrow +\infty$ and $\eps \to 0$ we obtain $w(x) \geq v^*(x)$. \\
To show the reverse inequality, we will prove that there exists an admissible strategy $\pi^*$ such that $w(x)= v(x,\pi^*)$. Let $(X_t^*,Z_t^*)$ be the solution of
\begin{equation}
X^*_{t} = \int_0^{t}(\mu-\alpha((1-X_{s^-})^+)) ds + \sigma B_{t} -  Z^*_{t}
\end{equation}
where,
\begin{equation}
Z^*_t=(x\ind_{\{x\le a\}}+(x-b)^+)1_{\{t=0^{-}\}} + \int_{0}^{t\wedge \tau_a^{-}}1_{\{X_s^*=b\}}dZ^*_s + a 1_{\{t \ge \tau_a\}}
\end{equation}

with
$$
\tau_a = \inf\{t \geq 0, X_{t-}^* \leq a\}
$$
whose existence is guaranteed by standard results on the Skorokhod problem (see for example Revuz and Yor \cite{ry:99}). The strategy $\pi^* = (Z_t^*)$ is admissible. Note also that $X^*_t$ is continuous on $[0,\tau_a^{-}]$. It is obvious that $v(x,\pi^*)=x=w(x)$ for $x\le a$. Now suppose $x > a$. Along the policy $\pi^*$, the liquidation time $\tau_0$ coincides with $\tau_a$ because $X^*_{\tau_a}=0$. Proceeding analogously as in the first part of the proof, we obtain

\begin{eqnarray*}
w(x)&=&\mathbb{E}_x\left[e^{-r(t\wedge \tau_0)} w(X^*_{t \wedge \tau_0})\right]+\mathbb{E}_x\left[\int_{0}^{t \wedge \tau_0^{-}}e^{-r s}w^{'}(X_s^*)dZ^*_s\right] + \mathbb{E}_x\left[ \ind_{t > \tau_0} e^{-r\tau_0}(w(X^*_{\tau_0-})-w(X^*_{\tau_0})) \right]\\
&=&\mathbb{E}_x\left[e^{-r(t\wedge \tau_0)} w(X^*_{t \wedge \tau_0})\right]+\mathbb{E}_x\left[\int_{0}^{t \wedge \tau_0^{-}}e^{-r s}w^{'}(b)dZ^*_s\right] +
\mathbb{E}_x\left[ \ind_{t > \tau_0}e^{-r\tau_0} a \right] \\
&=&\mathbb{E}_x\left[e^{-r(t\wedge \tau_0)} w(X^*_{t \wedge \tau_0})\right]+\mathbb{E}_x\left[\int_{0}^{t \wedge \tau_0}e^{-r s}dZ^*_s\right],
\end{eqnarray*}
where the last two equalities uses, $w(a)=a$ $w^{'}(b)=1$ and $(\Delta Z^*)_{\tau_0}=a$.
Now, because $w(0)=0$,
$$
\mathbb{E}_x\left[e^{-r(t\wedge \tau_0)} w(X^*_{t \wedge \tau_0})\right]=\mathbb{E}_x\left[e^{-rt} w(X^*_t)\ind_{t \le \tau_0}\right].
$$
Furthermore, because $w$ has at most linear growth and $\pi^*$ is admissible, we have
$$
\lim_{t\to\infty}\mathbb{E}_x\left[e^{-rt} w(X^*_t)\ind_{t \le \tau_0}\right]=0.
$$
Therefore, we have by letting $t$ tend to $+\infty$,
$$
w(x)=\mathbb{E}_x\left[\int_{0}^{\tau_0}e^{-r s}dZ^*_s\right]=v(x,\pi^*)
$$
which concludes the proof.
\end{pf}
\begin{rem} We notice that the proof remains valid when $a=0$ and $w'(0)$ is infinite by a standard localisation argument which will be the case in section 4.
\end{rem}

\subsection{Optimal Policies}

The verification theorem allows us to characterize the value function. The following theorem summarizes our findings.

\begin{theo}\label{MainResult}
Under Assumption \ref{hyp1} and \ref{hyp2}, the following holds:
\begin{itemize}
\item If $\mu \le r$, it is optimal to liquidate the firm, $v^*(x)=x$.
\item If $\mu \ge \alpha(1)$, the value of the firm is an increasing and concave function of the book value of equity. Any excess of cash above the threshold $b^*=\inf\{x>0, (v^{*})^{'}(x)=1\}$ is paid out to shareholders.(See Figure 1).
\item If $\mu < \alpha(1)$, the value of the firm is an increasing convex-concave function of the book value of equity. When the book value of equity is below the threshold $a=\sup\{x>0, v^*(x)=x\}$, it is optimal to liquidate. Any excess of cash above the threshold $b_a^*=\inf\{x>a, (v^{*})^{'}(x)=1\}$ is paid out to shareholders.(See Figure 2) 
\end{itemize} 
\end{theo}

It is interesting to compare our results with those obtained in the case of all equity financing. First, because the use of credit line is costly, it is optimal to wait that the cash reserves are depleted to draw on it. Moreover, there exists a target cash level above which it is optimal to pay out dividends. These two first findings are similar to the case of all equity financing. On the other hand, the marginal value of cash may not be monotonic in our case. Indeed, when the cost of the credit line is high, it becomes optimal for shareholders to terminate the lending relationship. This embedded option value makes the shareholder value locally convex in the neighborhood of the liquidation threshold $a$. The higher is the cost, measured by $\lambda$ in our simulation, the sooner is the strategic default or equivalently, the value function decreases, while the embedded exit option increases, with the cost of the credit line. The strategic default comes from the fact that the instantaneous firm’s profitability $\mu-\alpha(x)$ becomes negative for low value of equity capital. This is a key feature of our model that never happens when the firm is all-equity where the marginal value of cash at zero is the only statistic either to trigger the equity issuance or to liquidate.\\

\textit{Figure 1} plots some value functions, when $\mu \ge \alpha(1)$, using a linear function for $\alpha$, $\alpha(x)=\lambda x$ with different values of $\lambda$.

\begin{figure}[H]
\begin{center}
\includegraphics[width = 0.8\textwidth]{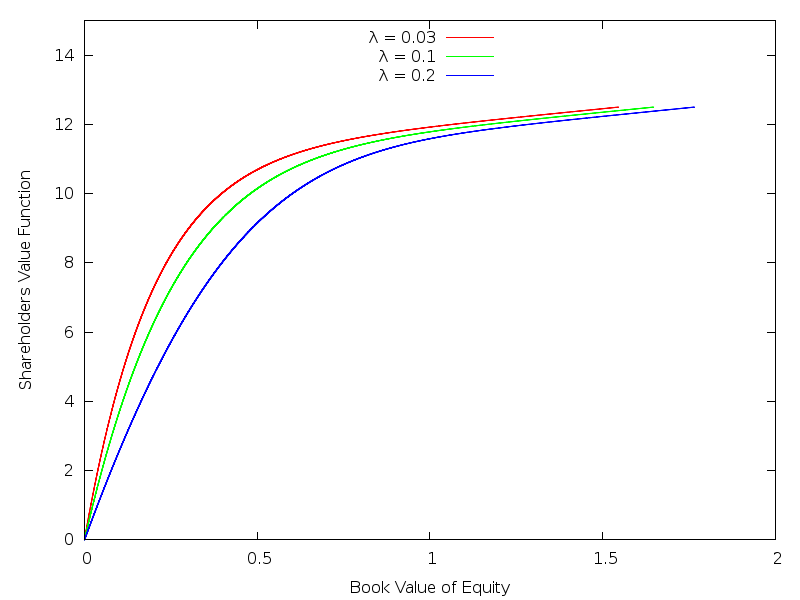}
\caption{Comparing  shareholders value functions with $\mu = 0.25$, $r = 0.02$, $\sigma = 0.3$ and $\mu \geq \alpha(1)$  for different values of $\lambda$ where  $\alpha(x)=\lambda x$.}
\end{center}
\end{figure}

\textit{Figure 2} plots some value functions, when $\alpha(1) > \mu$, using a linear function for $\alpha$, $\alpha(x) = \lambda x$ for different values of $\lambda$.

\begin{figure}[H]
\begin{center}
\includegraphics[width =  0.8\textwidth]{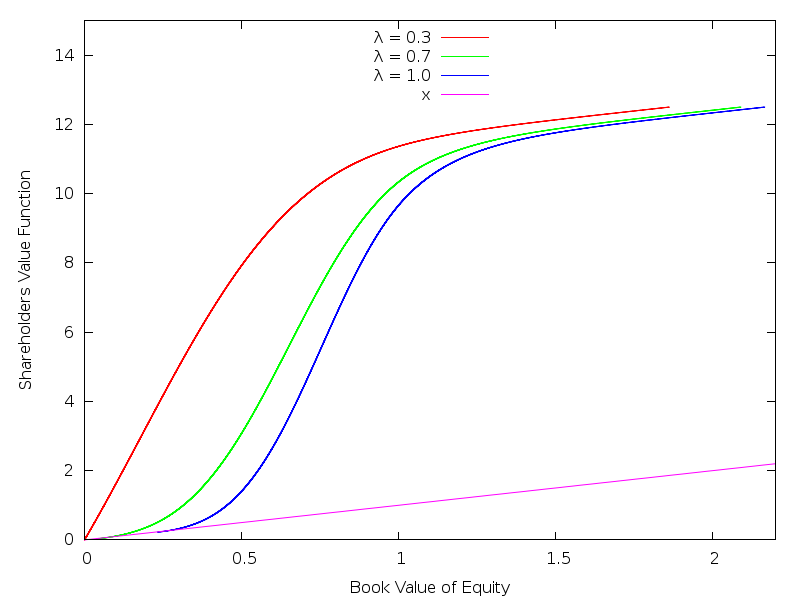}
\caption{Comparing shareholders value functions with $\mu = 0.25$, $r = 0.02$, $\sigma = 0.3$ and $\alpha(1) > \mu$ for different values of $\lambda$ where $\alpha(x) = \lambda x$.}
\end{center}
\end{figure}

Next section is devoted to the proof of Theorem \ref{MainResult}. The proof is based on an explicit construction of a smooth solution of the free boundary problem and necessitates a series of technical lemmas.

\section{Solving the free boundary problem}

The first statement of Theorem \ref{MainResult} comes from the fact that the function $\tilde w(x)=x$ satisfies Proposition \ref{propositionFB} when $\mu \le r$. To see this, we have to show that ${\cal L}\tilde w(x)$ is nonpositive for any $x\ge 0$. A straightforward computation gives 
$$
\hbox{ for } x>1, \quad{\cal L}\tilde w(x)=\mu-rx < \mu -r \leq 0,
$$
$$
\hbox{ for }x \le 1, \quad {\cal L}\tilde w(x)=\mu-\alpha(1-x)-rx.
$$
Using Equation \eqref{controlederiveealpha} of Assumption \ref{hyp1}, we observe that ${\cal L} \tilde w(x)$ is nondecreasing for $x\le 1$ and nonpositive at $x=1$ when $\mu \le r$.

Hereafter, we will assume that $\mu > r$ and focus on the existence of a function $w$ and a pair of constants $(a,b)$ satisfying Proposition \ref{propositionFB}. We will proceed in two steps. First we are going to establish some properties of the solutions of the differential equation ${\cal L} w = 0$. Second, we will consider two different cases- one where the productivity of the firm is always higher than the maximal interest payment $\alpha(1)\le \mu$, the other where the interest payment of the loan may exceed the productivity of the firm $\alpha(1) > \mu$.\\
Standard existence and uniqueness results for linear second-order differential equations imply that, for each $b$, the Cauchy problem :
\begin{equation} \label{cauchy-b}
\left\{
\begin{split}
r w(x) &= (\mu - \alpha((1-x)^+))w^{'}(x) + \frac{\sigma^2}{2}w^{''}(x)\\
w^{'}(b) &=1 \\
w^{''}(b) &=0 \\
\end{split}
\right.
\end{equation}
has a unique solution $w_b$ over $[0,b]$. By construction, this solution satifies $w_b(b) = \frac{\mu - \alpha((1-b)^+)}{r}$. Extending $w_b$ linearly to $[b,\infty[$ as $w_b(x) = x - b + \frac{\mu - \alpha((1-b)^+)}{r}$, for $x \geq b$
yields a twice continuously differentiable function over $[0,\infty[$, which is still denoted by $w_b$. 

\subsection{Properties of the solution to the Cauchy Problem}
We will establish a serie of preliminary results of the smooth solution $w_b$ of (\ref{cauchy-b}).

\begin{lem}\label{lem1}
Assume $b > 1$. If $w_b(0) = 0$ then $w_b$ is increasing and thus positive. 
\end{lem}
\begin{pf}
Because $w_b(0)=0$, $w_b(b)=\frac{\mu}{r}$ and ${\cal L}w_b=0$, the maximum principle implies $w_b > 0$ on $(0,+\infty)$.
Let us define 
$$
c=\inf\{ x >0, w_b^{'}(x)=0\}
$$
If $c=0$ then $w_b(0)=w^{'}_b(0)=w^{''}_b(0)=0$. By unicity of the Cauchy problem, this would imply $w_b=0$ which contradicts $w_b(b)=\frac{\mu}{r}$. Thus, $c>0$. If $c < b$, we would have $w_b(c)>0$, $w_b^{'}(c)=0$ and $w_b^{''}(c) \le 0$ and thus ${\cal L} w_b(c) <0$ which is a contradiction. 
Therefore $w^{'}_b$ is always positive.
\end{pf}

\begin{lem}\label{lem2}
Assume $b > 1$. We have $w_b^{'} > 1$ and $w^{''}_b < 0$ on $[1, b[$. 
\end{lem}
\begin{pf}
Because $w_b$ is smooth on $]1,b]$, we differentiate Equation (\ref{cauchy-b})  to obtain,
\[
w_b^{'''}(b) = \frac{2 r}{\sigma^2} > 0.
\]
As $w_b^{''}(b) = 0$ and $w_b^{'}(b) = 1$, it follows that $w_b^{''} < 0$, and thus $w_b^{'} > 1$ over some interval $]b-\epsilon, b[$, where $\epsilon > 0$. Now suppose by way of contradiction that $w_b'(x) \leq 1$ for some $x \in [1,b-\epsilon]$ and let $\tilde{x} = \sup \{x \in [1,b-\epsilon], w^{'}_b(x) \leq 1\}$. Then $w_b^{'}(\tilde{x})=1$ and $w_b'(x) > 1$ for $x \in ]\tilde{x},b[$, so that $w_b(b) - w_b(x) > b - x$ for all $x \in ]\tilde{x}, b[$. Because $w_b(b) = \frac{\mu}{r}$, this implies that for all $x \in ]\tilde{x}, b[$,
\[
w_b^{''}(x) = \frac{2}{\sigma^2}[r w_b(x) - \mu w_b^{'}(x)] <  \frac{2}{\sigma^2}[r (x - b + w_b(b)) - \mu] = \frac{2}{\sigma^2}r (x - b) < 0
\]
which contradicts $w'_b(b) = w'_b(\tilde{x}) = 1$. Therefore $w'_b > 1$ over $[1,b[$. Furthermore, using Lemma \ref{lem1}, 
\begin{eqnarray*}
w_b^{''}(x)&=& \frac{2}{\sigma^2}[r w_b(x) - \mu w_b^{'}(x)]\\ 
&<&\frac{2}{\sigma^2}[r w_b(x) - \mu ]\\
& <& \frac{2}{\sigma^2}[r w_b(b) - \mu ]\\
& =& 0.
\end{eqnarray*}
\end{pf}\\
The next result gives a sufficient condition on $b$ to ensure the concavity of $w_b$ on $(0,b)$.
\begin{coro}\label{coro1}
Assume $b \geq \frac{\alpha(1)}{r}$ and $ \mu \ge \alpha(1)$, we have $w'_b > 1$ and $w_b'' < 0$ over $]0,b[$. 
\end{coro}
\begin{pf}
Proceeding analogously as in the proof of Lemma \ref{lem2}, we define $\tilde{x} = \sup \{x \in [0,b-\epsilon], w'_b(x) \leq 1\}$ such that $w_b'(\tilde{x})=1$ and $w_b'(x) > 1$ for $x \in ]\tilde{x},b[$, so that $w_b(b) - w_b(x) > b - x$ for all $x \in ]\tilde{x}, b[$. Because $b > \frac{\alpha(1)}{r} > 1$, $w_b(b) = \frac{\mu}{r}$, we have
\[
\begin{split}
w_b^{''}(x) &= \frac{2}{\sigma^2}[r w_b(x) - (\mu-\alpha((1-x)^+)) w_b'(x)]\\
&<  \frac{2}{\sigma^2}[r (x - b + w_b(b)) - (\mu-\alpha((1-x)^+)]\\
&< \frac{2}{\sigma^2}[r (x - b) + \alpha((1-x)^+)].\\
\end{split}
\]
Denote by $g$ the function
\[
g(x) =  \frac{2}{\sigma^2}[r (x - b) + \alpha(1-x)], \qquad x \in [0,1[. 
\]
We have $g'(x) =  \frac{2}{\sigma^2}[r - \alpha'(1-x)]<0$ by Assumption \ref{hyp1}. Because $g(0) = \frac{2}{\sigma^2}[-r b + \alpha(1)] \leq 0$ if $b \geq \frac{\alpha(1)}{r}$, we have $w_b^{''}(x) < 0$ for $x \in ]0,1]$ which contradicts $w_b^{'}(\tilde{x}) = 1$ and $w'_b(1) > 1$ by Lemma \ref{lem2}. Therefore $w'_b > 1$ over $[0,1[$, from which it follows  $w''_b < 0$ and $w_b$ is concave on $]0,1[$. Because Lemma \ref{lem2} gives the concavity of $w_b$ on $[1,b[$, we conclude.
\end{pf}
The next proposition establishes some results about the regularity of the function $b \rightarrow w_b(y)$ for a fixed $ y\in [0,1[$.
\begin{lem}\label{lem3}
For any $ y\in [0,1[$, $b \rightarrow w_b(y)$ is an increasing function of b over $[y,1]$ and strictly decreasing over $]1, +\infty[$.
\end{lem}
\begin{pf} Consider the solutions $H^y_0$ and $H^y_1$ to the linear second-order differential equation ${\cal L}H = 0$ over $[y,\infty[$ characterized by the initial conditions $H^y_0(y) = 1$, $(H^y_0)'(y) = 0$, $H^y_1(y) = 0$, $(H^y_1)'(y) = 1$. We first show that $(H^y_0)'$ and $(H^y_1)'$ are strictly positive on $]y,\infty[$. Because $H^y_0(y) = 1$ and $(H^y_0)'(y) = 0$, one has $(H_0^y)''(y) = \frac{2r}{\sigma^2} > 0$, such that $(H^y_0)'(x) > 0$ over some interval $]y,y+\epsilon[$ where $\epsilon > 0$. Now suppose by way of contradiction that $\tilde{x} = \inf\{x \geq y+\epsilon, (H^y_0)'(x) \leq 0\} < \infty$. Then $(H^y_0)'(\tilde{x})=0$ and $(H^y_0)''(\tilde{x}) \leq 0$. Because ${\cal L}H^y_0 = 0$, it follows that $H^y_0(\tilde{x}) \leq 0$, which is impossible because $H^y_0(y) = 1$ and $H^y_0$ is strictly increasing over $[y,\tilde{x}]$. Thus $(H^y_0)' > 0$ over $]y;\infty[$, as claimed. The proof for $H^y_1$ is similar, and is therefore omitted.\\ Next, let $W_{H^y_0,H^y_1}=H^y_0(H^y_1)' - H^y_1(H^y_0)'$ be the Wronskian of $H^y_0$ and $H^y_1$. One has $W_{H^y_0,H^y_1}(y)=1$ and 
\[
\begin{split}
\forall x \geq y, \qquad W'_{H^y_0,H^y_1}(x) = &H^y_0(x)(H^y_1)''(x) - H^y_1(x)(H^y_0)''(x)\\
= &\frac{2}{\sigma^2}[H^y_0(x)(r H^y_1(x)-(\mu - \alpha((1-x)^+))(H^y_1)'(x))\\
&- H^y_1(x)(r H^y_0(x) -(\mu - \alpha((1-x)^+))(H^y_0)'(x))]\\
= &-\frac{2[\mu-\alpha((1-x)^+)]}{\sigma^2} W_{H^y_0,H^y_1}(x).\\
\end{split}
\]
Because $\alpha$ is integrable, the Abel's identity follows by integration:
\[
\forall x \geq y, \qquad W_{H^y_0,H^y_1}(x) = \exp\left[\frac{2}{\sigma^2} \left(-\mu (x-y) + \int_y^x \alpha ((1-u)^+)du\right)\right] .
\]
Because $W_{H^y_0,H^y_1} > 0$, $H^y_0$ and $H^y_1$ are linearly independent. As a result of this, $(H^y_0,H^y_1)$ is a basis of the two-dimensional space of solutions to the equation ${\cal L}H = 0$.
It follows in particular that for each $b > 0$, on can represent $w_b$ as :
\[
\forall x \in [y, b], \qquad w_b(x) = w_b(y) H^y_0(x) + w_b'(y)H^y_1(x).
\]
Using the boundary conditions $w_b(b)=\frac{\mu - \alpha((1-b)^+)}{r}$ and $w'_b(b)=1$, on can solve for $w_b(y)$ as follows:
\[
w_b(y) = \frac{(H^y_1)'(b)\frac{\mu - \alpha((1-b)^+)}{r}-H^y_1(b)}{W_{H^y_0,H^y_1}(b)}.
\]
Using the derivative of the Wronskian along with the fact that $H^y_1$ is solution to ${\cal L}H = 0$, it is easy to verify that:
\[
\begin{split}
\forall b \in [y,1[, \frac{dw_b(y)}{db} &= \frac{(H^y_1)'(b)\left(\frac{\alpha'(1-b)^+}{r}-1\right)}{W_{H^y_0,H^y_1}(b)}\\
\forall b \in ]1,\infty[, \frac{dw_b(y)}{db} &= \frac{-(H^y_1)'(b)}{W_{H^y_0,H^y_1}(b)}.\\
\end{split}
\]
So $w_b(y)$ is an increasing function of b over $[y,1]$ and strictly decreasing over $]1,\infty[$. 
\end{pf}

\begin{coro}\label{coro2}
If $b_2 > b_1 > 1$, then  $w_{b_2} < w_{b_1}$.
\end{coro}
\begin{pf}
Let us define $W = w_{b_1} - w_{b_2}$. Clearly, $W>0$ on $[b_2,+\infty[$. Moreover, we have ${\cal L}W = 0$ on $[0, b_1]$ and $W(0) > 0$ by Lemma \ref{lem3}.
 Moreover, $w_{b_1}(b_1) = w_{b_2}(b_2)$ and $w_{b_2}(b_2) > w_{b_2}(b_1)$ by Lemma \ref{lem2}.
 Therefore, the maximum principle implies $w_{b_2} < w_{b_1}$ on $[0, b_1]$. Finally, $w_{b_2}$ is concave and $w'_{b_2}(b_2) = 1$ therefore for $b_1 \le x \le b_2$,
\begin{eqnarray*}
w_{b_2}(x)&\le& w_{b_2}(b_2) +x-b_2\\
&=&\frac{\mu }{r}+x-b_2\\
&<&\frac{\mu }{r}+x-b_1\\
&=& w_{b_1}(x).
\end{eqnarray*}
\end{pf}

\subsection{Existence of a solution to the free boundary problem}

We are now in a position to characterize the value function and determine the optimal dividend policy. Two cases have to be considered: when the profitability of the firm is always higher than the maximal interest payment ($\mu \geq \alpha(1)$) and when the interest payment exceeds the profitability of the firm ($\mu < \alpha(1)$).

\subsubsection{Case: $\mu \geq \alpha(1)$}

The next lemma establishes the existence of a solution $w_{b^*}$ to the Cauchy problem (\ref{cauchy-b}) such that $w_{b^*}(0) = 0$.

\begin{lem} \label{existence_betoile}
There exists $ b^* \in ]1, \frac{\mu}{r}[$ such that the solution to (\ref{cauchy-b}) satisfies $ w_{b^*}(0) = 0$.
\end{lem}
\begin{pf}
Because $\mu \geq \alpha(1)$, we know from Corollary \ref{coro1} that $w_{\frac{\mu}{r}}$ is a concave function on $[0,\frac{\mu}{r}]$.
Moreover, because $\mu > r$, $w_{\frac{\mu}{r}}(\frac{\mu}{r})=\frac{\mu}{r}$. Because $w_{\frac{\mu}{r}}$ is strictly concave over $]0,\frac{\mu}{r}[$ with $w_{\frac{\mu}{r}}(\frac{\mu}{r})=\frac{\mu}{r}$ and $w^{'}_{\frac{\mu}{r}}=1$, $w_{\frac{\mu}{r}}(x) \le x$ for all $x \in ]0,\frac{\mu}{r}[$. In particular, $w_{\frac{\mu}{r}}(0) < 0$.\\ 
Moreover, we have :
\[
w_0(0) =\frac{\mu - \alpha(1)}{r} \geq 0.
\]
Therefore, Lemma \ref{lem3} implies $w_1(0) > 0$. Finally by continuity there is some $b^* \in ]1,\frac{\mu}{r}[$ such that $w_{b^*}(0) = 0$ which concludes the proof.
\end{pf}\\
The next lemma establishes the concavity of $w_{b^*}$.

\begin{lem}\label{lem4-concavityofwb*}
The function $w_{b^*}$ is concave on $[0,b^*]$
\end{lem}
\begin{pf}
Because $b^* > 1$, Lemma \ref{lem2} implies that $w_{b^*}$ is concave on $[1,b^*]$ thus $w''_{b^*}(1) \leq 0$.\\ For $x < 1$, we differentiate the differential equation satisfied by $w_{b^*}$ to get,
\begin{equation} \label{equaDeriv}
\frac{\sigma^2}{2}w'''_{b^*}(x) + (\mu - \alpha(1-x))w''_{b^*}(x) + (\alpha'(1-x) - r)w'_{b^*}(x) = 0.
\end{equation}
Because $w_{b^*}(0) = 0$ we have $w''_{b^*}(0) = -\frac{2}{\sigma^2}(\mu - \alpha(1))w'_{b^*}(0) \leq 0$ .\\
Now, suppose by a way of contradiction that $w''_{b^*} > 0$ on some subinterval of $[0,1]$. Because $w^{''}_{b^*}$ is continuous and nonpositive at the boundaries of $[0,1]$, there is some $c$ such that $w'''_{b^*}(c) = 0$ and $w''_{b^*}(c) > 0$. But, this implies
\[
w'_{b^*}(c) = -\frac{(\mu - \alpha(1-c))w''_{b^*}(c)}{\alpha'(1-c) - r} < 0
\]
which is a contradiction with Lemma \ref{lem1}.
\end{pf}

\begin{pro}
If $\mu \geq \alpha(1)$, $w_{b^*}$ is the solution of the control problem (9).
\end{pro}
\begin{pf}
Because $w_{b^*}$ is concave on $[0,b^*]$ and $w'(b^*) = 1$,  $w' \geq 1$ on $[0,b^*]$. Therefore we have a twice continuously differentiable concave function $w_{b^*}$ and a pair of constants $(a,b)=(0,b^*)$ satisfying the assumptions of Proposition \ref{propositionFB} and thus $w_{b^*} = v^*$.
\end{pf}\\

When the maximal interest payment is lower than the firm profitability, the value function is concave. This illustrates the shareholders' fear to liquidate a profitable firm. In particular, the shareholders value is a decreasing function of the volatility. 

\subsubsection{Case: $\mu < \alpha(1)$}

We first show that, for all $y \in [0,1[$, there exists $b_y$ such that $w_{b_y}$ is the solution of the Cauchy Problem \eqref{cauchy-b} with $w_{b_y}(y) = y$. 

\begin{lem} \label{lem5}
For all $y \in [0,1[$, we have $w_1(y)>y$.
\end{lem}
\begin{pf}
Because $\alpha$ is continuous with $\alpha(0) = 0$ and $\mu > r$, there exists
$\epsilon$ such that $w_{1-\epsilon}(1 - \epsilon) = \left(\frac{\mu - \alpha(\epsilon)}{r}\right) > 1$. Differentiating Equation \eqref{cauchy-b}, we observe
\[
w'''_{1-\epsilon}(1-\epsilon) = \frac{2}{\sigma^2}(r - \alpha'(\epsilon)) < 0
\] 
using Equation \eqref{controlederiveealpha}. Therefore $w_{1-\epsilon}$ is convex in a left neighborhood of $1-\varepsilon$. If $w_{1-\epsilon}$ is convex on $(0,1-\varepsilon)$ then $w_{1-\epsilon}(x) \ge x - ( 1 - \epsilon ) + \frac{\mu - \alpha(\epsilon)}{r} > x$ for $\varepsilon$ small enough and the result is proved.\\
If $w_{1-\epsilon}$ is not convex on $(0,1-\varepsilon)$ then it will exist some $\bar{x} < 1-\varepsilon$ such that $w^{''}_{1-\epsilon}(\bar{x})=0$, $w^{'''}_{1-\epsilon}(\bar{x})>0$ and $w_{1-\epsilon}$ convex on $]\bar{x},1-\epsilon]$. Differentiating Equation \eqref{cauchy-b} at $\bar{x}$ gives $w^{'}_{1-\epsilon}(\bar{x})<0$. Therefore $w_{1-\epsilon}$ is nonincreasing in a neighborhood of $\bar{x}$. Assume by a way of contradiction that $w_{1-\epsilon}$ is increasing at some point $\hat x \in [0,\bar x[$. This would imply the existence of $\tilde{x}<\bar{x}$ such that $w^{'}_{1-\epsilon}(\tilde{x}) = 0$, $w^{''}_{1-\epsilon}(\tilde{x}) < 0$ and $w_{1-\epsilon}(\tilde{x}) > 0$ which contradicts Equation \eqref{cauchy-b}. 
Therefore $w_{1-\epsilon}$ is decreasing on $(0,\bar{x})$ and convex on $(\bar{x},1-\varepsilon)$ which implies that $w_{1-\epsilon}(x)>x$ for all $x \le 1-\varepsilon$. 
To conclude, for any $y<1$, we can find $\varepsilon$ small enough to have 
$w_{1-\epsilon}(y)>y$ which can be extended to $w_1(y)>y$ by Lemma \ref{lem3}.
\end{pf}
\begin{coro} \label{coro3}
For all $y \in [0,1[$, there is an unique $b_y \in ]1, 1+ \frac{\mu}{r}[$ such that $w_{b_y}(y) = y$.
\end{coro}
\begin{pf}
By Lemma \ref{lem1}, $w_{ 1+\frac{\mu}{r}}$ is concave on $]1,1+\frac{\mu}{r}[$, thus $w_{ 1+\frac{\mu}{r}}(1) < \frac{\mu}{r} +(1-(1+\frac{\mu}{r})) =0$.
Suppose that there exists $c$ in $[0,1[$ such that $w_{ 1+\frac{\mu}{r}}(c)>0$, then there exists $\tilde x \in ]c,1[$ such that $w_{ 1+\frac{\mu}{r}}(\tilde x)<0$, $w^{'}_{ 1+\frac{\mu}{r}}(\tilde x)=0$,  $w^{''}_{ 1+\frac{\mu}{r}}(\tilde x)>0$ yielding to the standard contradiction with the maximum principle. We thus have $w_{1+\frac{\mu}{r}}(y) < y$ for all $y \in [0,1+\frac{\mu}{r}]$. Using Lemma \ref{lem5} and the continuity of the function $b \rightarrow w_b(y)$, it exists for all $y<1$ a threshold $b_y\in ]1,1+ \frac{\mu}{r}[ $ such that $w_{b_y}(y) = y$. The uniqueness of $b_y$ comes from Corollary \ref{coro2}.
\end{pf}\\

We will now study the behavior of the first derivative of $w_{b_y}$. 
\begin{lem}\label{lem6}
There exists $\epsilon > 0$ such that $w^{'}_{b_{1-\varepsilon}}(1-\varepsilon) \ge 1  $ and $b_{1-\epsilon} < \frac{\mu}{r}$.
\end{lem}
\begin{pf}
Because $\alpha(0) = 0$ and $\mu > r$, it exists $\eta > 0$ such that 
\begin{eqnarray}
\forall x \in [1-\eta,1], \alpha(1-x) + r x - \mu < 0.
\label{alphaIneg}
\end{eqnarray}
 Moreover $w_{\frac{\mu}{r}}$ is strictly concave on $[1,\frac{\mu}{r}[$ by Lemma \ref{lem2} and thus 
\begin{eqnarray*}
w_{\frac{\mu}{r}}(1)& \le& w_{\frac{\mu}{r}}(\frac{\mu}{r})+(1-\frac{\mu}{r})w^{'}_{\frac{\mu}{r}}(\frac{\mu}{r})\\
&=&1.
\end{eqnarray*}
Because by Lemma \ref{lem2}, we have $w'_{\frac{\mu}{r}} > 1$ on $[1, \frac{\mu}{r}[$, there exists $\nu > 0$ such that $\forall x \in [1-\nu,1], w_{\frac{\mu}{r}}(x) < x$. Let $\varepsilon=\min(\eta,\nu)$. By Corollary \ref{coro3}, it exists $b_{1-\epsilon} \in ]1,1+ \frac{\mu}{r}[$ such that $w_{b_{1-\epsilon}}(1 - \epsilon) = 1 - \epsilon$. We have $w_{b_{1-\epsilon}}(1 - \epsilon) > w_{\frac{\mu}{r}}(1-\epsilon)$ and then $b_{1-\epsilon} < \frac{\mu}{r}$ by Corollary \ref{coro2}.\\
Let us consider the function $W(x) = w_{b_{1-\epsilon}}(x) - x$, we have $W(1-\epsilon) = 0$, $W(b_{1-\epsilon}) = \frac{\mu}{r} - b_{1-\epsilon} > 0$. Moreover, $W$ is solution 
\begin{equation}\label{equadiffpourW}
(\mu - \alpha((1-x)^{+})W'(x) + \frac{\sigma^2}{2}W''(x)-r W(x) = \alpha((1-x)^{+}) + r x - \mu.
\end{equation}
On $[ 1-\epsilon,1]$, the second member of Equation \eqref{equadiffpourW} is negative due to Equation \eqref{alphaIneg}.
On $[1,b_{1-\epsilon}]$ , it is equal to $r x - \mu$  which is negative because $b_{1-\epsilon}< \frac{\mu}{r}$.
Assume by a way of contradiction that there is some $x \in [1-\epsilon, b_{1-\epsilon}]$ such that $W(x) < 0$, then it would exist $\tilde{x} \in [1-\epsilon, b_{1-\epsilon}]$ such that $W(\tilde{x}) < 0, W'(\tilde{x}) = 0$ and $W''(\tilde{x}) > 0$ which is in contradiction with Equation \eqref{equadiffpourW}.
Hence, $W$ is a positive function on $[1-\epsilon, b_{1-\epsilon}]$ with $W(1-\epsilon) = 0$ which implies
 $w'_{b_{1-\epsilon}}(1 - \epsilon) \ge 1$.
\end{pf}

\begin{lem}\label{lem7}
When $\mu<\alpha(1)$, $w_{b_0}$ is a convex-concave function.
\end{lem}
\begin{pf}
According to Corollary \ref{coro3}, there exists $b_0 \in  ]1, 1+\frac{\mu}{r}[$ such that  $w_{b_0}(0) = 0$ and by Lemma \ref{lem1}, $w'_{b_0} > 0$ on $(0,b_0)$. Using Equation \eqref{cauchy-b}, we thus have $w^{''}_{b_0}(0) > 0$  implying that $w_{b_0}$ is strictly convex on a right neighborhood of $0$. Because $b_0 > 1$, Lemma \ref{lem2} implies $w_{b_0}''(x) < 0$ on $[1,b_0[$. If there is more than one change in the concavity of $w_{b_0}$, it will exist $\bar{x} \in [0,1[$ such that $w'''_{b_0}(\bar{x}) > 0$, $w''_{b_0}(\bar{x}) = 0$ and $w_{b_0}'(\bar{x}) \geq 0$ yielding the standard contradiction.
\end{pf}

\begin{pro} \label{solutionbzero}
If $\mu < \alpha(1)$ and $w'_{b_0}(0) \geq 1$, $w_{b_0}$ is the shareholders value function \eqref{value}
\end{pro}
\begin{pf}
It is straightforward to see that the function $w_{b_0}$ satisfies Proposition \ref{propositionFB} when $w'_{b_0}(0) \geq 1$.
\end{pf}

\noindent Now, we will consider the case $w'_{b_0}(0) < 1$. 

\begin{lem}
If $w'_{b_0}(0) < 1$, it exists $a \in ]0,1[$ such that $w_{b_a}(a) = a$ and $w'_{b_a}(a) = 1$. 
\end{lem}
\begin{pf}
Let $\phi(x)=w^{'}_{b_x}(x)$. By assumption, we have $\phi(0) < 1$ and by Lemma \ref{lem6}, $\phi(1-\varepsilon) > 1$. By continuity of $\phi$, there exists $a \in ]0,1[$ such that $w_{b_a}'(a) = 1$. By definition, the function $w_{b_a}$ satisfies $w_{b_a}(a) = a$. 
\end{pf}

\begin{lem}\label{lem8}
$w_{b_a}$ is a convex-concave function on $[a,b_a]$.
\end{lem}
\begin{pf}
First, we show that $w_{b_a}$ is increasing on $[a,b_a]$.
 Because  $w_{b_a}'(a)=1$, we can define $\tilde{x} = \min\{x > a, w_{b_a}'(x) \leq 0 \}$.  If $\tilde x \le b_a$, we will have $w'_{b_a}(\tilde{x}) = 0$, $w_{b_a}(\tilde{x}) > 0$ and $w''_{b_a}(\tilde{x}) \le 0$ yielding the standard contradiction.
According to Lemma \ref{lem1}, we have $w''_{b_a}(x) < 0$ over $[1,b_a[$ because $b_a > 1$. Proceeding analogously as in the proof of Lemma \ref{lem7}, we prove that $w_{b_a}$ is a convex-concave function because it cannot change of concavity twice.
\end{pf}

\begin{lem} \label{lem9}
We have $w_{b_a}> 1$ on $(a,b_a)$ with $b_a < \frac{\mu}{r}$.
\end{lem}
\begin{pf}
According to Lemma \ref{lem8}, $w_{b_a}$ is convex-concave with $w_{b_a}'(a) = 1$ and $w_{b_a}'(b_a) = 1$, therefore $\forall x \in ]a,b_a[, w_{b_a}'(x) > 1$. As a consequence, $w_{b_a}(x) > x$ on $]a, b_a]$ and in particular $w_{b_a}(1) > 1$. Remembering that $w_{\frac{\mu}{r}}(1) < 1$ and using Corollary \ref{coro2}, we have $b_a < \frac{\mu}{r}$. 
\end{pf}

\begin{pro}
If $w'_{b_0}(0) < 1$, the function
$$
w(x)=\left\{ \begin{array}{cc} x& \hbox{ for }x\le a\\
w_{b_a}(x)& \hbox{ for } a \le x \le b_a\\
x-b_a+\frac{\mu}{r}& \hbox{ for } x \ge b_a
\end{array}
\right.
$$
is the shareholders value function \eqref{value}.
\end{pro}
\begin{pf}
it is straightforward to check that $w$ satisfies Proposition \ref{propositionFB}.
\end{pf}

\section{The Investment Model}

In this section, we enrich the model to allow variable investment in the productive assets. We will assume a decreasing-returns-to-scale technology by introducing an increasing concave function $\beta$ with $\lim_{x \rightarrow \infty} \beta(x) = \bar{\beta}$ that impacts the dynamic of the book value of equity as follows: 
\begin{equation}\label{dynamique}
\left\{
\begin{split}
dX_t &= \beta(K_t)(\mu dt + \sigma dW_t) - \alpha((K_t-X_t)^+)dt - \gamma |dI_t| - dZ_t\\
dK_t &= dI_t = dI_t^+ - dI_t^-\\
\end{split}
\right.
\end{equation}
where $I_t^+$ (resp. $I_t^-$) is the cumulative capital invested (resp. disinvested) in the productive assets up to time $t$, $\gamma>0$ is an exogenous proportional cost of investment. Assumption \eqref{hyp2}  thus forces liquidation when the level of outstanding debt reaches the sum of the liquidation value of the productive assets and the liquid assets, $(1-\gamma)K_t + M_t$. 
The goal of the management is to maximize over the admissible strategies $\pi = (Z_t,I_t)_{t\geq0}$ the risk-neutral shareholders value
\begin{equation}\label{fonc_obj}
V^*(x,k) = \sup_\pi\mathbb{E}_{x,k}\left(\int_0^{\tau_0} e^{-rt}dZ_t\right)
\end{equation}
where
$$
\tau_0=\inf\{t\geq 0, L_t \geq (1-\gamma)K_t + M_t\} = \inf\{t \geq 0, X_t \leq \gamma K_t\}.
$$
By definition, we have
\begin{equation}\label{cond_init}
\forall k \geq 0, V^*(\gamma k,k) = 0.
\end{equation}

\subsection{Dynamic programming and free boundary problem}
In order to derive a classical analytic characterization of $V^*$ in terms of a free boundary problem, we rely on the dynamic programming principle as follows\\
\noindent {\it Dynamic Programming Principle:} For any $(x,k) \in S$ where $S=\{(x,k)\in \mathbb{R}_+^2 ,\, x \geq \gamma k\}$, we have
\begin{equation}\label{dynamicprogramming}
V^*(x,k)=\sup_\pi\mathbb{E}\left(\int_0^{\theta} e^{-rt}dZ_t+e^{-r\theta}V^*(X_\theta,K_\theta)\right)
\end{equation}
where $\theta$ is any stopping time.\\
Take the suboptimal control $\pi$ which consists in investing only at time $t=0$ a certain amount $h$. Then, according to the dynamic programming principle, we have with $\theta = 0^+$,
\[
V^*(x,k) \geq V^*(X_{0^+},K_{0^+}) = V^*(x- \gamma h, k + h).
\]
So,
\[
V^*(x,k) - V^*(x-\gamma h,k) + V^*(x-\gamma h, k) - V^*(x-\gamma h, k+ h) \geq 0.
\]
Dividing by $h$, we have
\[
\gamma \frac{V^*(x,k) - V^*(x-\gamma h,k)}{\gamma h} - \frac{ V^*(x-\gamma h, k+ h) - V^*(x-\gamma h, k) }{h} \geq 0.
\]
If $V^*$ were smooth enough, we can let $h$ tend to $0$ to obtain
\[
\gamma \frac{\partial V^*}{\partial x}- \frac{\partial V^*}{\partial k} \geq 0.
\]
Likewise, we can prove that
\[
\begin{split}
\gamma \frac{\partial V^*}{\partial x} + \frac{\partial V^*}{\partial k} \geq 0 \\
\frac{\partial V^*}{\partial x} - 1 \geq 0\\
\end{split}
\]
and
\[
- \mathcal{L}_k V^* \geq 0
\]
where $\mathcal{L}_k$ is the second order differential operator
\begin{equation}\label{op_diff}
\mathcal{L}_kw = \big(\beta (k)\mu - \alpha ((k-x)^+)\big) \frac{\partial w}{\partial x} + \frac{\sigma^2 \beta(k)^2}{2}\frac{\partial^2 w}{\partial x^2} - r w.
\end{equation}
The aim of this section is to characterize via the dynamic programming principle the shareholders value as the unique continuous viscosity solution to the free boundary problem in order to use a numerical procedure to describe the optimal policies.
\begin{equation}\label{fb_viscosite}
F(x,k,V^*,DV^*,D^2V^*)=0
\end{equation}
where
\[
F(x,k,w,Dw,D^2w) = \min\left(-\mathcal{L}_kw,\frac{\partial w}{\partial x} - 1,\gamma \frac{\partial w}{\partial x}- \frac{\partial w}{\partial k},\gamma \frac{\partial w}{\partial x}+ \frac{\partial w}{\partial k} \right).
\]

We will first establish the continuity of the shareholders value function which relies on some preliminary well-known results about hitting times we prove below for sake of completeness.

\begin{lem}\label{lem_temps_d_arret}
Let $a<b$ and $(x_n)_{n\geq0}$ a sequence of real numbers such that $\lim_{n\rightarrow +\infty} x_n = b$ and $ \min_n x_n > a$. Let $(X_t^n)_{n\geq0}$ the solution of the stochastic differential equation
\[
\left\{
\begin{split}
dX_t^n &= \mu_n(X^n_t) dt + \sigma_n dW_t\\
X_0^n &= x_n\\
\end{split}
\right.
\]
where $\mu_n$ and $\sigma_n$ satisfy the standard global Lipschitz and linear growth conditions. Moreover, $(\sigma_n)_{n \geq 0}$ are strictly positive real numbers converging to $\sigma > 0$ and $(\mu_n)_{n\geq 0}$ is a sequence of bounded functions converging uniformly to $\mu$.
Let us define $T_n = \inf\{t \geq 0, X_t^n = a\}$ and $\theta_n = \inf\{t \geq 0, X_t^n = b\}$. We have
\[
\lim_{n\rightarrow +\infty}\mathbb{P}(\theta_n < T_n) = 1.
\]
\end{lem}
\begin{pf}
Let us define the functions $U_n, F_n : I \rightarrow \mathbb{R}$, on some bounded interval I containing $(a,b)$ as
\[
U_n(y) = \int_0^y\mu_n(z+x_n)dz, \qquad F_n(y) = \int_0^ye^{-\frac{2U_n(z)}{\sigma_n^2}}dz.
\]
Because $(\mu_n)_{n\geq 0}$ converges uniformly to $\mu$, we note that  $(F_n,U_n)_{n\geq 0}$ converges uniformly to {\bf $(F,U)$} where
\[
F(y) = \int_0^ye^{-\frac{2U(z)}{\sigma^2}}dz
\]
and
\[
U(y) = \int_0^y\mu(z+b)dz.
\]
Let $Y_t^n = X^n_t - x_n$, $M^n_t = F_n(Y^n_t)$ and $\tau_n = \inf\{t \geq 0, Y^n_t \notin ]a_n,b_n[\}$ with $a_n = a - x_n$ and $b_n = b - x_n$. We first show that $\tau_n$ is integrable. Because $F_n$ is the scale function of the process $Y_t^n$,  $M^n_t$ is a local martingale with quadratic variation
\[
<M^n>_t = \int_0^t\sigma_n^2 e^{-\frac{4U_n(Y^n_s)}{\sigma_n^2}}ds.
\]
Because 
$$
E(<M^n>_{t\wedge _{\tau_n}}) \le \sigma_n^2t  \exp\left(-\frac{4}{\sigma_n^2}\min_{y\in[a_n,b_n]}U_n(y)\right) < +\infty
$$
the processes $(M^n_{t\wedge \tau_n})_{t \geq 0}$ and $((M^n_{t\wedge\tau_n})^2 - <M^n>_{t\wedge \tau_n})_{t \geq 0}$ are both martingales. By Optional sampling theorem
\[
\mathbb{E}[(M^n_{t\wedge\tau_n})^2 - <M^n>_{t\wedge \tau_n}] = 0 
\] 
which implies
\[
\mathbb{E}\left[\int_0^t1_{[0,\tau_n]}(s)\sigma_n^2e^{-\frac{4U_n(Y^n_s)}{\sigma_n^2}}ds\right] = \mathbb{E}[F_n^2(Y^n_{t\wedge \tau_n})]
\]
and
\[
\sigma_n^2 \exp\left(-\frac{4}{\sigma_n^2} \max_{y\in[a_n,b_n]}U_n(y)\right) \mathbb{E}[t\wedge \tau_n] \leq \max_{y \in [a_n,b_n]}F_n^2(y)
\]
thus there is a constant $K_n > 0$ such that
\[
\forall t \geq 0, \mathbb{E}[t\wedge \tau_n] \leq K_n.
\]
We conclude by dominated convergence that $\tau_n$ is integrable. The martingale property implies
\[
\mathbb{E}[F_n(Y^n_{t \wedge \tau_n})] = 0
\]
which yields
\[
\mathbb{E}[F_n(Y^n_{\tau_n})] = 0,
\]
by dominated convergence because
\[
\forall t \geq 0, |F_n(Y^n_{t\wedge\tau_n})| \leq \max_{y\in[a_n,b_n]}|F_n(y)|.
\]
This is equivalent to
\[
F_n(a_n)(1-p(a_n,b_n)) + F_n(b_n)p(a_n,b_n) = 0
\]
with $p(a_n,b_n) = \mathbb{P}(Y^n_{\tau_n} = b_n)$.
Hence,
\[
p(a_n,b_n) = \frac{-F_n(a_n)}{F_n(b_n) - F_n(a_n)} .
\]
Moreover,
\[
\begin{split}
\mathbb{P}(\theta_n < T_n) &= \mathbb{P}(X^n_{\tau_n} = b)\\
&= \mathbb{P}(Y^n_{\tau_n} = b - x_n)\\
&= p(a_n,b_n).\\
\end{split}
\]
Using the uniform convergence of $F_n$, we have
\[
\begin{split}
\lim_{n\rightarrow +\infty} \mathbb{P}(\theta_n < T_n) =& \lim_{n\rightarrow +\infty}p(a_n,b_n)\\
=& \frac{-F(a-b)}{F(0)-F(a-b)}\\
=& 1 .
\end{split}
\]
\end{pf}

\begin{lem}\label{lem_tranf_laplace}
Let $a<b$ and $(x_n)_{n\geq0}$ a sequence of real numbers such that $\lim_{n\rightarrow +\infty} x_n = b$ and $ \min_n x_n > a$. Let $(X_t^n)_{n\geq0}$ the solution to
\[
\left\{
\begin{split}
dX_t^n &= \mu_n(X^n_t) dt + \sigma_n dW_t\\
X_0^n &= x_n\\
\end{split}
\right.
\]
with the same assumptions as in Lemma \ref{lem_temps_d_arret}. There exist constants $A_n$ and $B_n$ such that 
\begin{equation}\label{ineg_2_3}
\exp\left(-\frac{b-x_n}{\sigma_n^2}(\sqrt{A_n^2+2r \sigma_n^2} - A_n)\right)\leq \mathbb{E}[e^{-r \theta_n}] \leq \exp\left(-\frac{b-x_n}{\sigma_n^2}(\sqrt{B_n^2+2r \sigma_n^2} - B_n)\right).
\end{equation}
\end{lem}

\begin{pf} Because $\mu_n$ are bounded functions, there are two constants $A_n$ and $B_n$ such that $A_n \le \mu_n(x) \le B_n$ for all $a<x<b$. We define $\tilde{X}^n_t=x_n+A_n t+ \sigma_n W_t$.  By comparison, we have $\tilde{X}^n_t \leq X^n_t$ and $\theta_n \leq \tilde{\theta}_n$, with $\tilde{\theta}_n = \inf\{t\geq 0, \tilde{X}^n_t = b\}$.
But the Laplace transform of $\tilde{\theta_n}$ is explicit and given by
\[
\mathbb{E}[e^{-r \tilde{\theta}_n}] = \exp\left(-\frac{b-x_n}{\sigma_n^2}(\sqrt{A_n^2+2r \sigma_n^2} - A_n)\right)
\]
which gives the left inequality of \eqref{ineg_2_3}. The proof is similar for the right inequality introducing $\bar{X_t}^n=x_n+B_n t+ \sigma_n W_t$.
\end{pf}

\begin{pro}\label{prop_cont}
The shareholders value function is jointly continuous.
\end{pro}

\begin{pf}
Let $(x,k)\in S$ and let us consider $(x_n,k_n)$ a sequence in $S$ converging to $(x,k)$. Therefore, $\{(x_n - \gamma |k - k_n|, k), (x-\gamma|k-k_n|, k_n)\} \in S^2$ for $n$ large enough.
We consider the following two strategies that are admissible for $n$ large enough:
\begin{itemize}
\item Strategy $\pi^1_n$: start from $(x,k)$, invest if $k_n - k>0$(or disinvest if $k_n-k<0$) and do nothing up to the minimum between the liquidation time and the hitting time of $(x_n,k_n)$. Denote $(X^{\pi^1_n}_t,K^{\pi^1_n}_t)_{t\geq0}$ the control process associated to strategy $\pi^1_n$.
\item Strategy $\pi^2_n$: start from $(x_n,k_n)$, invest if $k_n - k<0$(or disinvest if $k_n-k>0$) and do nothing up to the minimum between the liquidation time and the hitting time of $(x,k)$. Denote $(X^{\pi^2_n}_t,K^{\pi^2_n}_t)_{t\geq0}$ the control process associated to strategy $\pi^2_n$.
\end{itemize}
To fix the idea, assume $k_n>k$. The strategy $\pi_1$ makes the process $(X,K)$ jump from $(x,k)$ to $(x-\gamma(k_n-k),k_n)$.
\begin{center}
\includegraphics[scale=0.5]{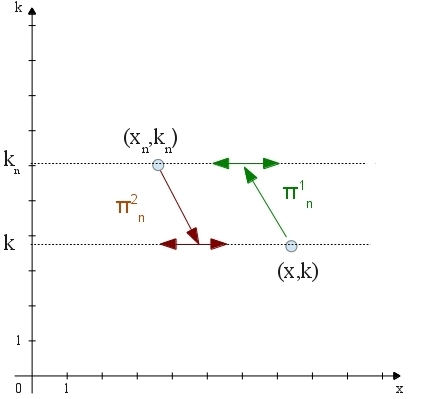}
\end{center}
Define
\[
\theta^1_n = \inf\{t \geq 0, (X^{\pi^1_n}_t,K^{\pi^1_n}_t) = (x_n,k_n)\},
\]
\[
\theta^2_n = \inf\{t \geq 0, (X^{\pi^2_n}_t,K^{\pi^2_n}_t) = (x,k)\},
\]
\[
T^1_n = \inf\{t \geq 0, X^{\pi^1_n,x}_t \leq \gamma K^{\pi^1_n,k}_t\}
\]
and
\[
T^2_n = \inf\{t \geq 0, X^{\pi^2_n,x_n}_t \leq \gamma K^{\pi^2_n,k_n}_t\}.
\]
Dynamic programming principle and $V^*(X_{T^1_n},K_{T_n^1})=0$ on $T^1_n \leq \theta^1_n$ yield
\begin{equation}\label{ineg_droite}
\begin{split}
V^*(x,k) &\geq \mathbb{E}\left[\int_0^{\theta^1_n \wedge T^1_n}e^{-r t}dZ^{\pi^1_n}_t + e^{-r(\theta^1_n \wedge T^1_n)} 1_{\{\theta^1_n < T^1_n\}}V^*(X_{\theta^1_n},K_{\theta_n^1})\Big) \right]\\
&\geq \mathbb{E}\left[e^{-r\theta^1_n} 1_{\{\theta^1_n < T^1_n\}}V^*(x_n,k_n)\right]\\
&\geq \left(\mathbb{E}\big(e^{-r\theta^1_n}\big) - \mathbb{E}\big(e^{-r\theta^1_n}1_{\{\theta^1_n \geq T^1_n\}}\big)\right)V^*(x_n,k_n)\\
&\geq \left(\mathbb{E}\big(e^{-r\theta^1_n}\big) - \mathbb{P}\big(\theta^1_n \geq T^1_n\big)\right)V^*(x_n,k_n).\\
\end{split}
\end{equation}
On the other hand, using $V^*(X_{T^2_n},K_{T_n^2})=0$ on $T^2_n \leq \theta^2_n$
\begin{equation}\label{ineg_gauche}
\begin{split}
V^*(x_n,k_n) &\geq \mathbb{E}\left[\int_0^{\theta^2_n \wedge T^2_n}e^{-r t}dZ^{\pi^2_n}_t + e^{-r(\theta^2_n \wedge T^2_n)} 1_{\{\theta^2_n < T^2_n\}}V^*(X_{\theta^2_n},K_{\theta_n^2})\Big) \right]\\
&\geq \mathbb{E}\left[e^{-r\theta^2_n} 1_{\{\theta^2_n < T^2_n\}}V^*(x,k)\right]\\
&\geq \left(\mathbb{E}\big(e^{-r\theta^2_n}\big) - \mathbb{E}\big(e^{-r\theta^2_n}1_{\{\theta^2_n \geq T^2_n\}}\big)\right)V^*(x,k)\\
&\geq \left(\mathbb{E}\big(e^{-r\theta^2_n}\big) - \mathbb{P}\big(\theta^2_n \geq T^2_n\big)\right)V^*(x,k).\\
\end{split}
\end{equation}
The convergence of $(x_n,k_n)$ implies
\[
\lim_{n\rightarrow +\infty}(x_n - \gamma |k-k_n|,k) = (x,k)
\]
from which we deduce using Lemma \ref{lem_temps_d_arret} that
\begin{equation}\label{lim1}
\lim_{n\rightarrow +\infty}\mathbb{P}(\theta_n^1 \geq T_n^1) = 0
\end{equation}
and
\begin{equation}\label{lim2}
\lim_{n\rightarrow +\infty}\mathbb{P}(\theta_n^2 \geq T_n^2) = 0.
\end{equation}
Let $\mu_n(X^n_t) = \beta(k_n)\mu - \alpha((k_n - X^n_t)^{+})$ and $\sigma_n = \beta(k_n)\sigma$. The function $\mu_n$ is bounded by
\[
\begin{split}
A_n &= \beta(k_n)\mu - \alpha(k_n)\\
B_n &= \beta(k_n)\mu\\
\end{split}
\]
thus, according to Lemma \ref{lem_tranf_laplace}
\[
\exp\left(-\frac{\kappa^n}{\sigma_n^2}(\sqrt{A_n^2+2r \sigma_n^2} - A_n)\right)  \leq \mathbb{E}[e^{-r \theta^1_n}] \leq \exp\left(-\frac{\kappa^n}{\sigma_n^2}(\sqrt{B_n^2+2r \sigma_n^2} - B_n)\right)
\]
with 
$\kappa^n = x-x^n +\gamma |k^n-k|$.\\
Letting $n$ tend to $+\infty$ and using

$$
\begin{array}{ll}
\lim_{n\rightarrow +\infty} A_n &= \beta(k)\mu - \alpha(k)\\
\lim_{n\rightarrow +\infty} B_n &= \beta(k)\mu\\
\lim_{n\rightarrow +\infty} \sigma_n &= \beta(k)\sigma\\
\end{array}
$$

we obtain
\begin{equation}\label{lim3}
\lim_{n\rightarrow +\infty}\mathbb{E}(e^{- r \theta_n^1}) = \lim_{n\rightarrow +\infty}\mathbb{E}(e^{- r \theta_n^2}) =1.
\end{equation}
Finally, we have from \eqref{ineg_droite} and \eqref{ineg_gauche}, 
$$
V^*(x,k)\ge \limsup_n V^*(x_n,k_n) \ge \liminf_n V^*(x_n,k_n) \ge V^*(x,k),
$$
which proves the continuity of $V^*$.
\end{pf} 

We are now in a position to characterize the shareholders value in terms of viscosity solution of the free boundary problem \eqref{fb_viscosite}.
\begin{theo}\label{charac_viscosity}
The shareholders value $V^*$ is the unique continuous viscosity solution to \eqref{fb_viscosite} on $S$ with linear growth.
\end{theo} 

\begin{pf}
The proof is postponed to the Appendix
\end{pf}

The main interest of Theorem \ref{charac_viscosity} is to guarantee that the standard numerical procedure to solve HJB free boundary problems proposed in \cite {fl:07} will converge to the shareholders value function. We obtain the following description of the control regions (Figure 3). Our numerical analysis demonstrates that \begin{itemize}
\item unlike \cite{bcw:11}, there exists an optimal level of productive assets (top of the yellow region) and thus an objective measure of managerial overinvestment in our context. This is clearly due to the decreasing-returns-to-scale assumption.
\item constrained firms with low cash reserves, that is when equity capital is close to productive asset size, and low equity capital will rather disinvest to offset cash-flows shortfalls.
\item constrained firms with low cash reserves and high equity capital will first draw on the credit line to offset cash-flows shortfalls. 
\item the credit line is never used to invest. 
\end{itemize}

\begin{figure}[H]
\begin{center}
\includegraphics[width =  \textwidth]{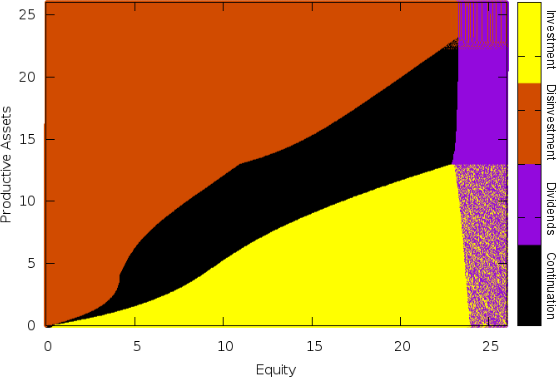}
\caption{Optimal control with $\mu = 0.25$, $r = 0.02$, $\sigma = 0.3$, $\lambda = 0.08$, $\beta_{max} = 20$, $\bar{\beta} = 10$ and an investment cost $\gamma = 5e^{-4}$.}
\end{center}
\end{figure}

While the numerical results give the above insights about the optimal policies, we have not been able to prove rigorously the shape of the optimal control regions. Nonetheless, making the strong assumption that there is no transaction cost $\gamma=0$ allows us to fully describe the control regions and gives us reasons to believe in Figure 3. This is the object of our last subsection.

\subsection{Absence of Investment cost}

Using a verification procedure analogous to section 3, we characterize the value function and the optimal policies in terms of a free boundary problem. The following proposition proved in the Appendix summarizes our findings.

\begin{pro} \label{Resultgamma=0}
When there is no cost of investment/disinvestment, $\gamma=0$, the following holds:

\begin{itemize}

\item If $\mu \beta^{'}(0) \le r$ then it is optimal to liquidate the firm thus $v^*(x)=x$.

\item If $\alpha^{'}(0)> \mu \beta^{'}(0) >r$ and $\sigma^2\beta'(0) \ge \frac{\mu}{(1-\delta)}$, the shareholders value is an increasing and concave function of the book value of equity. Any excess of cash above the threshold $b^*=\inf\{x>0, (v^{*})^{'}(x)=1\}$ is paid out to shareholders (see Figure 4). The optimal size of the productive asset is characterized by a deterministic function of equity capital (see Figure 5) given by
\[
\begin{split}
\forall 0 \leq x \leq a, k(x) &= \beta^{-1}\left[\frac{\mu x}{\sigma^2(1-\delta)}\right]\\
\forall x \geq a, k(x) &= x.\\
\end{split}
\]

where $a$ is the unique nonzero solution of the equation
\begin{equation}
\label{equationpoura}
\sigma^2(1-\delta) \beta(a) = \mu a.
\end{equation}
with
\begin{equation}
\label{delta}
\delta=\frac{2r \sigma^2}{\mu^2 + 2r \sigma^2}
\end{equation}

\item If $\alpha^{'}(0)> \mu \beta^{'}(0) >r$ and $\sigma^2\beta'(0) < \frac{\mu}{(1-\delta)}$, the shareholders value is an increasing and concave function of the book value of equity (see Figure 6). Any excess of cash above the threshold $b^*=\inf\{x>0, (v^{*})^{'}(x)=1\}$ is paid out to shareholders. Moreover, all the cash reserves are invested in the productive assets.

\end{itemize} 

\end{pro}

The above proposition has two interesting implications.
\begin{itemize} 
\item When the volatility of earnings is low  $\sigma^2\beta'(0) < \frac{\mu}{(1-\delta)}$, it is optimal to invest all the cash reserves in the productive assets and use it as a complementary substitute for cash which is better off than using a costly credit line.
\item Nonetheless, when the volatility of earnings is high, productive assets are not a perfect substitute of cash because it implies a high risk of bankrupcy when the book value of equity is low.
\end{itemize}

\textit{Figure 4} plots the shareholders value functions with $\alpha^{'}(0) > \mu \beta'(0)$  and $\sigma^2 \beta'(0) \geq \frac{\mu}{(1-\delta)}$ for different values of $\beta'(0)$ using :
\begin{itemize}
\item a linear function for $\alpha$, $\alpha(x) = \lambda x$.
\item an exponential function for $\beta$, $\beta(x) = \beta_{max}\left(1 - e^{\frac{-\beta'(0)}{\beta_{max}}x}\right)$.
\end{itemize}

\begin{figure}[H]
\begin{center}
\includegraphics[width = \textwidth]{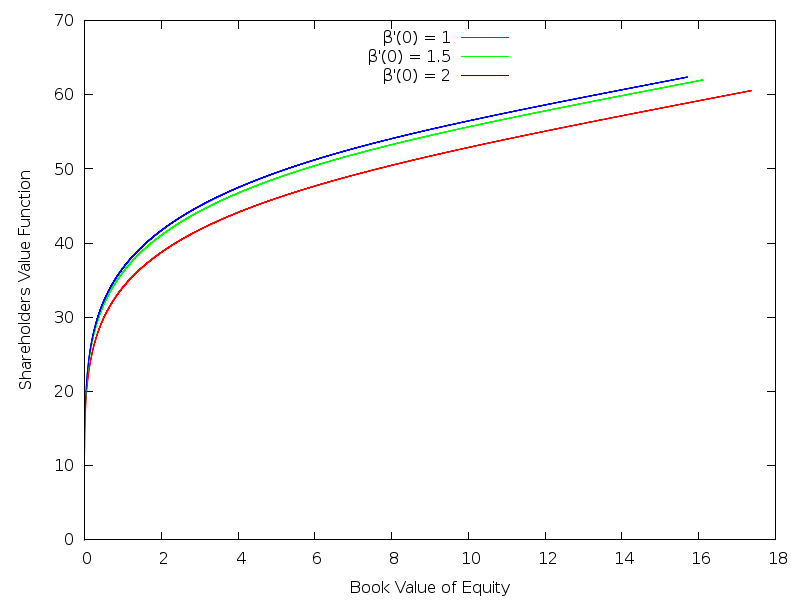}
\caption{Comparing  shareholders value functions with $\mu = 0.25$, $r = 0.02$, $\sigma = 0.6$, $\lambda = 0.8$, $\beta_{max} = 5$, for different values of $\beta'(0)$ (case  $\sigma^2 \beta'(0) \geq \frac{\mu}{(1-\delta)}$).}
\end{center}
\end{figure}

\noindent \textit{Figure 5}  plots the optimal level of productive assets for different values of $\sigma$. It shows that, for a given level of the book value of equity, the investment level in productive assets is a decreasing function of the volatility.

\begin{figure}[H]
\begin{center}
\includegraphics[width = \textwidth]{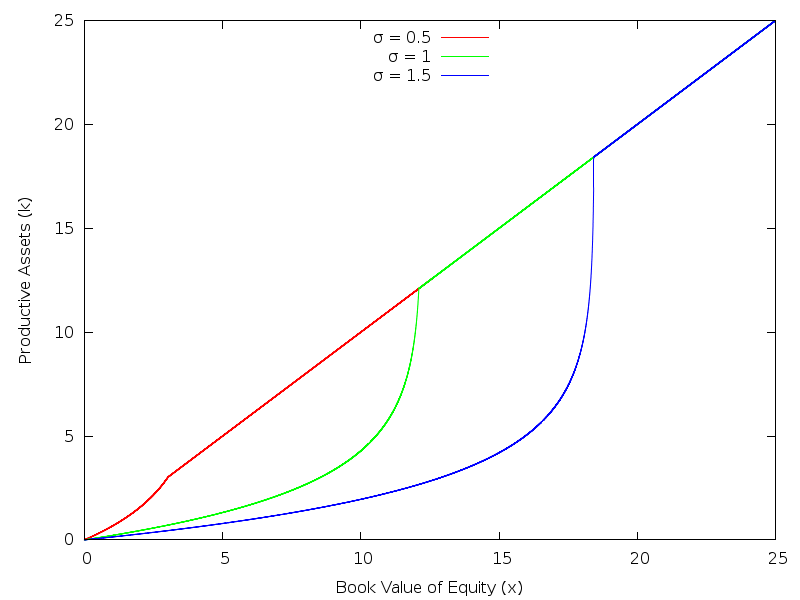}
\caption{Comparing  optimal level of productive assets with $\mu = 0.25$, $r = 0.02$, $\lambda = 0.8$, $\beta_{max} = 5$, $\beta'(0)=2$ for different values of $\sigma$.}
\end{center}
\end{figure}

\textit{Figure 6} plots the shareholders value functions when $\alpha'(0) > \mu \beta'(0)$  and $\sigma^2 \beta'(0) \leq \frac{\mu}{(1-\delta)}$ for different values of $\beta'(0)$ using :
\begin{itemize}
\item a linear function for $\alpha$, $\alpha(x) = \lambda x$.
\item an exponential function for $\beta$, $\beta(x) = \beta_{max}\left(1 - e^{\frac{-\beta'(0)}{\beta_{max}}x}\right)$.
\end{itemize}

\begin{figure}[H]
\begin{center}
\includegraphics[width = \textwidth]{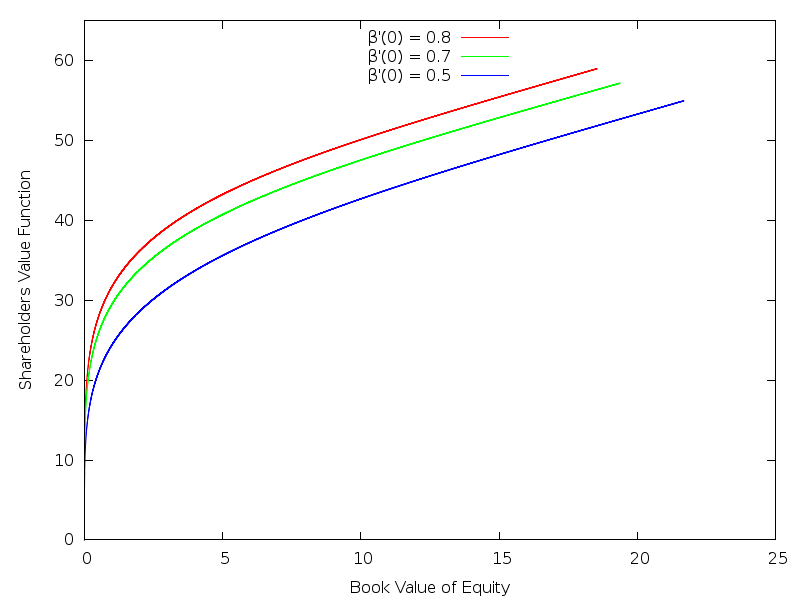}
\caption{Comparing  shareholders value functions with $\mu = 0.25$, $r = 0.02$, $\sigma = 0.6$, $\lambda = 0.8$, $\beta_{max} = 5$, for different values of $\beta'(0)$ (case $\sigma^2 \beta'(0) \leq \frac{\mu}{(1-\delta)}$).}
\end{center}
\end{figure}

\section{Appendix}
\subsection{Proof of Theorem \ref{charac_viscosity}}
{\it Supersolution property}. Let $(\bar{x},\bar{k}) \in S$ and $\varphi \in C^2(\R_+^2)$ s.t. $(\bar{x},\bar{k})$ is a
minimum of $V^*-\varphi$ in a neighborhood $B_\eps(\bar{x},\bar{k})$ of $(\bar{x},\bar{k})$ with $\eps$ small enough to ensure $B_\eps \subset S$ and $V^*(\bar{x},\bar{k}) =
\varphi(\bar{x},\bar{k})$.\\
First, let us consider  the admissible control
$\hat{\pi} = (\hat{Z}, \hat{I})$ where the shareholders decide
to never invest or disinvest, while the dividend policy is defined by
$\hat Z_t$ $=$ $\eta$ for $t$ $\geq$ $0$, with $0 \leq \eta \leq
\eps$. Define the exit time $\tau_\eps$  $=$ $\inf\{t \geq 0,
(X_t^{\bar x},K_t^{\bar k}) \notin \overline{B}_\eps(\bar{x},\bar{k}) \}$. We notice
that $\tau_\eps < \tau_0$ for $\eps$ small enough. From the dynamic programming principle, we have 
\begin{eqnarray}
\varphi(\bar{x},\bar{k}) = V^*(\bar{x},\bar{k}) &\geq&
\E \left[\int_0^{\tau_\eps \wedge h} e^{-r t}d\hat{Z}_t +
e^{-r(\tau_\eps \wedge h)} V^*(X_{\tau_\eps \wedge h}^{\bar x},K_{\tau_\eps \wedge h}^{\bar k})  \right ]\nonumber\\
&\geq& \E \left[\int_0^{\tau_\eps \wedge h} e^{-r t}d\hat{Z}_t +
e^{-r(\tau_\eps \wedge h)} \varphi(X_{\tau_\eps \wedge h}^{\bar x},K_{\tau_\eps \wedge h}^{\bar k})  \right ].  \label{proofvis2}
\end{eqnarray}
Applying  It\^o's formula to the process $e^{-r t}\varphi(X_t^{\bar x},K_t^{\bar{k}})$ between $0$
and $\tau_\eps \wedge h$, and taking the expectation, we obtain
\begin{eqnarray}
\E \left[e^{-r (\tau_\eps \wedge h)} \varphi(X_{\tau_\eps \wedge h}^{\bar x},K_{\tau_\eps \wedge h}^{\bar k}) \right] &=& \varphi(\bar{x},\bar{k})
+ \E \left[\int_0^{\tau_\eps \wedge h} e^{-r t} {\cal L} \varphi (X_t^{\bar x},K_t^{\bar{k}}) dt \right] \nonumber\\
&& +\; \E \left[ \sum_{0\leq t \leq \tau_\eps \wedge h}  e^{-r t} [\varphi(X_t^{\bar x},K_t^{\bar{k}}) - \varphi(X_{t^-}^{\bar x},K_t^{\bar{k}})] \right].
\label{proofvis3}
\end{eqnarray}
Combining relations \reff{proofvis2} and \reff{proofvis3}, we have
\begin{eqnarray}
\E \left[\int_0^{\tau_\eps \wedge h} e^{-r t}(- {\cal L}) \varphi (X_t^{\bar x},K_t^{\bar{k}}) dt \right]
- \E \left[\int_0^{\tau_\eps \wedge h} e^{-r t}d\hat{Z}_t \right] && \nonumber\\
- \E \left[ \sum_{0\leq t \leq \tau_\eps \wedge h}  e^{-r t} [\varphi(X_t^{\bar x},K_t^{\bar{k}}) - \varphi(X_{t^-}^{\bar x},K_t^{\bar{k}})] \right] &\geq& 0.  \label{proofvis4}
\end{eqnarray}

\begin{itemize}
\item[$\star$]  Take first $\eta = 0$. We then observe that $X$ is continuous on $[0, \tau_\eps \wedge h]$ and only the first term of the relation
\reff{proofvis4} is non zero. By dividing the above inequality by $h$ with $h \rightarrow 0$, we conclude that
$
- {\cal L} \varphi(\bar{x},\bar{k}) \geq 0.
$

\item[$\star$] Take now  $\eta > 0$ in \reff{proofvis4}. We see  that $\hat{Z}$ jumps only at $t = 0$ with size $\eta$, so that
$$
\E \left[\int_0^{\tau_\eps \wedge h} e^{-r t}( - {\cal L} \varphi)(X_t^{\bar x},K_t^{\bar{k}}) dt \right]
- \eta -(\varphi(\bar{x} - \eta,\bar{k}) - \varphi(\bar{x},\bar{k})) \geq 0.
$$
By sending  $h \rightarrow 0$, and then dividing by $\eta$ and letting $\eta \rightarrow 0$, we obtain
$$
\frac{\partial \varphi}{\partial x}(\bar{x},\bar{k}) - 1 \geq 0.
$$
\end{itemize}
Second, let us consider the admissible control
$\bar{\pi} = (\bar{Z}, \bar{I})$ where the shareholders decide
to never payout dividends, while the investment/disinvestment policy is defined by $\bar I_t=\eta \in \R$ for $t$ $\geq$ $0$, with $0<|\eta| \leq
\eps$. Define again the exit time $\tau_\eps$  $=$ $\inf\{t \geq 0,
(X_t^{\bar x},K_t^{\bar k}) \notin \overline{B}_\eps(\bar{x},\bar{k}) \}$.\\
Proceeding analogously as in the first part and observing that $\bar I$ jumps only at $t=0$, thus
$$
\E \left[\int_0^{\tau_\eps \wedge h} e^{-r t}( - {\cal L} \varphi)(X_t^{\bar x},K_t^{\bar{k}}) dt \right]
- (\varphi(\bar{x} - \gamma|\eta|,\bar{k}+\eta) - \varphi(\bar{x},\bar{k})) \geq 0.
$$
Assuming first $\eta>0$, by sending  $h \rightarrow 0$, and then dividing by $\eta$ and letting $\eta \rightarrow 0$, we obtain
$$
\gamma \frac{\partial \varphi}{\partial x}(\bar x,\bar k)-\frac{\partial \varphi}{\partial k}(\bar x,\bar k) \ge 0.
$$
When $\eta <0$, we get in the same manner
$$
\gamma \frac{\partial \varphi}{\partial x}(\bar x,\bar k)+\frac{\partial \varphi}{\partial k}(\bar x,\bar k) \ge 0.
$$
This proves the required supersolution property.\newline

{\it Subsolution Property:} We prove the subsolution property by contradiction. Suppose that the claim is not true. Then, there exists $(\bar{x},\bar{k}) \in S$ and a neighbourhood $B_\eps(\bar{x},\bar{k})$ of $\bar{x},\bar{k}$, included in $S$ for $\eps$ small enough, a $C^2$ function $\varphi$ with $(\varphi - V^*)(\bar{x},\bar{k})= 0$ and $\varphi \geq V^*$ on $B_\eps(\bar{x},\bar{k})$,
and $\eta > 0$, s.t. for all $(x,k) \in
B_\eps(\bar{x},\bar{k})$ we have
\begin{eqnarray} \label{proofvis5}
 - \Lc \varphi(x,k) &> \eta,\\
\frac{\partial \varphi}{\partial x}(x,k) - 1 &> \eta, \label{proofvis6} \\
(\gamma \frac{\partial \varphi}{\partial x}-\frac{\partial \varphi}{\partial k})(x,k) &> \eta. \label{proofvis7}\\
(\gamma \frac{\partial \varphi}{\partial x}+\frac{\partial \varphi}{\partial k})(x,k) &> \eta. \label{proofvis7bis}
\end{eqnarray}
For any admissible control $\pi$,
consider the exit time 
$\tau_\eps = \inf \{ t \geq 0, (X_t^{\bar{x}},K_t^{\bar{k}}) \notin B_\eps(\bar{x},\bar{k}) \}$ and  notice again that $\tau_\eps < \tau_0$. Applying  It\^o's formula to the process
 $e^{-r t}\varphi(X_t^{\bar{x}},K_t^{\bar{k}})$ between $0$
and $\tau_\eps^{-}$, we have

\beq
\mathbb{E}[e^{-r \tau_\eps^-}\varphi(X_{\tau_\eps^-},K_{\tau_\eps^-})] &=& \varphi(\bar{x},\bar{k}) - \mathbb{E}\left[\int_0^{\tau_\eps^-}e^{-ru}\mathcal{L}\varphi du\right]\\ 
&+& \mathbb{E}\left[\int_0^{\tau_\eps^-}e^{-r u}(-\gamma \frac{\partial \varphi}{\partial x} + \frac{\partial \varphi}{\partial k})dI_u^{c,+}\right]\\
&+& \mathbb{E}\left[\int_0^{\tau_\eps^-}e^{-r u}(-\gamma \frac{\partial \varphi}{\partial x} - \frac{\partial \varphi}{\partial k})dI_u^{c,-}\right]\\
&-& \mathbb{E}\left[\int_0^{\tau_\eps^-}e^{-r u}\frac{\partial \varphi}{\partial x}dZ^c_u\right]\\
&+& \mathbb{E}\left[\sum_{0 < s < \tau_\eps}e^{-r s}[\varphi(X_s,K_s) - \varphi(X_{s^-},K_{s^-})]\right] \label{proofvis8}
\enq
Using relations \eqref{proofvis5},\eqref{proofvis6},\eqref{proofvis7},\eqref{proofvis7bis}, we obtain

\beq
V^*(\bar{x},\bar{k}) &=& \varphi(\bar{x},\bar{k}) \nonumber\\
&\geq &  \eta \mathbb{E} \left[ \int_0^{\tau_\eps^-} e^{-r u} du \right] + \mathbb{E}[e^{-r \tau_\eps^-}\varphi(X_{\tau_\eps^-},K_{\tau_\eps^-})]\\
&+& \eta \mathbb{E}\left[\int_0^{\tau_\eps^-}e^{-r u}dI_u^{c,+}\right]\\
&+& \eta \mathbb{E}\left[\int_0^{\tau_\eps^-}e^{-r u}dI_u^{c,-}\right]\\
&+& (1+\eta)\mathbb{E}\left[\int_0^{\tau_\eps^-}e^{-r u}dZ^c_u\right]\\
&-& \mathbb{E}\left[\sum_{0 < s < \tau_\eps}e^{-r s}[\varphi(X_s,K_s) - \varphi(X_{s^-},K_{s^-})]\right]
\enq

Note that $\Delta X_s = -\Delta Z_s - \gamma ( \Delta I_s^+ + \Delta I_s^- )$, $\,\Delta K_s = \Delta I_s^+ - \Delta I_s^-$ and
by the Mean Value Theorem, there is some $\theta \in ]0,1[$ such that,
\beqs
\varphi(X_s,K_s) &-& \varphi(X_{s^-},K_{s^-}) = \frac{\partial \varphi}{\partial x}(X_{s^-} + \theta \Delta X_s, K_{s^-} + \theta \Delta K_s) \Delta X_s + \\ 
&  & \frac{\partial \varphi}{\partial k}(X_{s^-} + \theta \Delta X_s, K_{s^-} + \theta \Delta K_s)\Delta K_s\\
&=& \frac{\partial \varphi}{\partial x}(X_{s^-} + \theta \Delta X_s, K_{s^-} + \theta \Delta K_s)(-\Delta Z_s - \gamma ( \Delta I_s^+ + \Delta I_s^- ) ) \\
&+& \frac{\partial \varphi}{\partial k}(X_{s^-} + \theta \Delta X_s, K_{s^-} + \theta \Delta K_s)(\Delta I_s^+ - \Delta I_s^-)\\
&=& - \frac{\partial \varphi}{\partial x}(X_{s^-} + \theta \Delta X_s, K_{s^-} + \theta \Delta K_s)\Delta Z_s\\
&+& \left(- \gamma \frac{\partial \varphi}{\partial x}(X_{s^-} + \theta \Delta X_s, K_{s^-} + \theta \Delta K_s) + \frac{\partial \varphi}{\partial k}(X_{s^-} + \theta \Delta X_s, K_{s^-} + \theta \Delta K_s)\right) \Delta I_s^+\\
&+& \left(- \gamma \frac{\partial \varphi}{\partial x}(X_{s^-} - \theta \Delta X_s, K_{s^-} + \theta \Delta K_s) + \frac{\partial \varphi}{\partial k}(X_{s^-} + \theta \Delta X_s, K_{s^-} + \theta \Delta K_s)\right) \Delta I_s^-\\
\enqs

Because $(X_s+\theta \Delta X_s, K_s+\theta \Delta K_s) \in B_\eps (\bar{x},\bar{k})$, we use the  relations \eqref{proofvis6},\eqref{proofvis7},\eqref{proofvis7bis} again
$$
- (\varphi(X_s,K_s) - \varphi(X_{s^-},K_{s^-})) \geq (1+\eta) \Delta Z_s + \eta \Delta I_s^+ + \eta \Delta I_s^-
$$

Therefore,
\beqs
V^*(\bar{x},\bar{k}) &\geq & \mathbb{E}[e^{-r \tau_\eps^-}\varphi(X_{\tau_\eps^-},K_{\tau_\eps^-})] + \mathbb{E}\left[\int_0^{\tau_\eps^-}e^{-r u} dZ_u\right]\\
&+&  \eta \left( \mathbb{E} \left[ \int_0^{\tau_\eps^-} e^{-r u} du \right] + \mathbb{E}\left[\int_0^{\tau_\eps^-}e^{-ru}dI_u^+\right] + \mathbb{E}\left[\int_0^{\tau_\eps^-}e^{-r u}dI_u^-\right] + \mathbb{E}\left[\int_0^{\tau_\eps^-}e^{-r u}dZ_u\right] \right)
\enqs
Notice that while $(X_{\tau_\eps}^-, K_{\tau_\eps}^-) \in B_\eps(\bar{x},\bar{k})$, $(X_{\tau_\eps},K_{\tau_\eps})$ is either on the boundary $\partial B_\eps(\bar{x},\bar{k})$ or out of $\bar{B}_\eps(\bar{x},\bar{k})$. However, there is some random variable $\alpha$ valued in $[0,1]$ such that:
\[
\begin{split}
(X^{(\alpha)},K^{(\alpha)}) &= (X_{\tau_\eps^-},K_{\tau_\eps^-}) + \alpha (\Delta X_{\tau_\eps}, \Delta K_{\tau_\eps})\\
&= (X_{\tau_{\eps}^-},K_{\tau_{\eps}^-}) + \alpha (- \Delta Z_{\tau_\eps} - \gamma \Delta I^+_{\tau_\eps} - \gamma \Delta I^-_{\tau_\eps}, \Delta I^+_{\tau_\eps}- \Delta I^-_{\tau_\eps}) \in \partial B_\eps(\bar{x},\bar{k}).
\end{split}
\] 

Proceeding analogously as above, we show that
\[
\varphi(X^{(\alpha)},K^{(\alpha)}) - \varphi(X_{\tau_\eps^-},K_{\tau_\eps^-}) \leq - \alpha [(1+\eta) \Delta Z_{\tau_\eps} + \eta \Delta I_{\tau_\eps}^+ + \eta \Delta I_{\tau_\eps}^-].
\]
Observe that
\[
(X^{(\alpha)},K^{(\alpha)}) = (X_{\tau_\eps},K_{\tau_\eps}) + (1-\alpha)(\Delta Z_{\tau_\eps} + \gamma \Delta I^+_{\tau_\eps} + \gamma \Delta I^-_{\tau_\eps}, -\Delta I^+_{\tau_\eps} + \Delta I^-_{\tau_\eps}).
\]
Starting from $(X^{(\alpha)},K^{(\alpha)})$, the strategy that consists in investing $(1-\alpha)\Delta I_{\tau_\eps}^+$ or disinvesting $(1-\alpha)\Delta I_{\tau_\eps}^-$ depending on the sign of $K^{(\alpha)}-K_{\tau_\eps}$ and payout $(1-\alpha)\Delta Z_{\tau_\eps}$ as dividends leads to $(X_{\tau_\eps},K_{\tau_\eps})$ and therefore,
\[
V^*(X^{(\alpha)},K^{(\alpha)}) - V^*(X_{\tau_\eps},K_{\tau_\eps}) \geq (1- \alpha) \Delta Z_{\tau_\eps}.
\]

Using $\varphi(X^{(\alpha)},K^{(\alpha)}) \geq V^*(X^{(\alpha)},K^{(\alpha)})$, we deduce
\[
\varphi(X_{\tau_{\eps}^-},K_{\tau_{\eps}^-}) - V^*(X_{\tau_{\eps}}, K_{\tau_{\eps}}) \geq (1 + \alpha \eta) \Delta Z_{\tau_{\eps}} + \alpha \eta (\Delta I^+_{\tau_{\eps}} + \Delta I^-_{\tau_{\eps}}).
\]
Hence,
\beq \label{minDP}
V^*(\bar{x},\bar{k}) & \geq &  \eta \Big( \mathbb{E}\left[ \int_0^{\tau_{\eps}^-} e^{-r u} du \right] + \mathbb{E}\left[\int_0^{\tau_{\eps}^-}e^{-r u}dI_u^+\right] +\mathbb{E}\left[\int_0^{\tau_{\eps}^-}e^{-r u}dI_u^-\right] + \mathbb{E}\left[\int_0^{\tau_{\eps}^-}e^{-r u}dZ_u\right] \nonumber\\
&+& \mathbb{E}[e^{-r \tau_{\eps}}\alpha(\Delta Z_{\tau_{\eps}} + \gamma \Delta I_{\tau_{\eps}}^+ + \gamma \Delta I_{\tau_{\eps}}^-)]\Big)\nonumber\\
 &+& \mathbb{E}[e^{-r \tau_{\eps}}V^*(X_{\tau_{\eps}},K_{\tau_{\eps}})] + \mathbb{E}\left[\int_0^{\tau_{\eps}}e^{-r u}dZ_u\right]
\enq
We now claim there is $c_0 > 0$ such that for any admissible strategy
\beq \label{claim}
c_0 &\leq &\mathbb{E} \left[ \int_0^{\tau_{\eps}^-} e^{-r u} du + \int_0^{\tau_{\eps}^-}e^{-r u}dI_u^+ + \int_0^{\tau_{\eps}^-}e^{-r u}dI_u^- + \int_0^{\tau_{\eps}^-}e^{-r u}dZ_u\right]\\
&+& \mathbb{E}\left[e^{-r \tau_{\eps}}\alpha(\Delta Z_{\tau_{\eps}} + \gamma \Delta I_{\tau_{\eps}}^+ + \gamma \Delta I_{\tau_{\eps}}^-) \right] \nonumber
\enq
Let us consider the $C^2$ function, $\phi(x,k) = c_0[1-\frac{(x-\bar{x})^2}{\eps^2}]$ with,
\[
0 < c_0 \leq \min\left\{\frac{\eps}{2}, \frac{\eps}{2 \gamma}, \frac{1}{r}, \frac{\eps^2}{\sigma_n^2\bar{\beta}^2}, \frac{\eps}{2d_{max}}\right\}
\]
where
$$
d_{max} = \sup\left\{\frac{|\beta(k)\mu - \alpha((k-x)^+)|}{\eps}, (x,k) \in B_\eps(\bar{x},\bar{k})\right\}>0,
$$
satisfies
\[
\left\{
\begin{split}
&\phi(\bar{x},\bar{k}) = c_0\\
&\phi = 0, \qquad \text{ for } (x,k) \in \partial B_\eps\\ 
&\min\left\{1 - \mathcal{L}\phi, 1 -\gamma\frac{\partial \phi}{\partial x} + \frac{\partial \phi}{\partial k}, 1 -\gamma\frac{\partial \phi}{\partial x} - \frac{\partial \phi}{\partial k}, 1 - \frac{\partial \phi}{\partial x}\right\} \geq 0, \text{ pour } (x,k) \in B_\eps.\\
\end{split}
\right.
\]
Applying It\^o's formula, we have
\beq \label{eqforc0}
0 < c_0 &=& \phi(\bar{x},\bar{k}) \leq \mathbb{E}[e^{-r \tau_{\eps}^-}\phi(X_{\tau_{\eps}^-},K_{\tau_{\eps}^-})] + \mathbb{E}\left[\int_0^{\tau{\eps}^-}e^{-r u}du\right]\nonumber\\
 &+& \mathbb{E}\left[\int_0^{\tau_{\eps}^-}e^{-r u}dI^+_u\right]+ \mathbb{E}\left[\int_0^{\tau_{\eps}^-}e^{-r u}dI^-_u\right]+ \mathbb{E}\left[\int_0^{\tau_{\eps}^-}e^{-r u}dZ_u\right]
\enq
Noting that $\frac{\partial \phi}{\partial x} \leq 1$ and $\frac{\partial \phi}{\partial k}= 0$, we have
\[
\phi(X_{\tau_{\eps}^-},K_{\tau_{\eps}^-}) - \phi(X^{(\alpha)},K^{(\alpha)}) \leq (X_{\tau_{\eps}^-} - X^{(\alpha)}) = \alpha(\Delta Z_{\tau_{\eps}} + \gamma \Delta I^+_{\tau_{\eps}} + \gamma \Delta I^-_{\tau_{\eps}}).
\]
Plugging into \eqref{eqforc0} with $\phi(X^{(\alpha)},K^{(\alpha)})=0$, we obtain
\beqs
c_0 &\leq &\mathbb{E} \left[ \int_0^{\tau_{\eps}^-} e^{-r u} du + \int_0^{\tau_{\eps}^-}e^{-r u}dI_u^+ + \int_0^{\tau_{\eps}^-}e^{-r u}dI_u^- + \int_0^{\tau_{\eps}^-}e^{-r u}dZ_u\right]\\
&+& \mathbb{E}\left[e^{-r \tau_{\eps}}\alpha(\Delta Z_{\tau_{\eps}} + \gamma \Delta I_{\tau_{\eps}}^+ + \gamma \Delta I_{\tau_{\eps}}^-) \right]
\enqs
This proves the claim \eqref{claim}. Finally, by taking the supremum over $\pi$ and using the dynamic programming principle, \eqref{minDP} implies
$V^*(\bar{x},\bar{k}) \geq V^*(\bar{x},\bar{k}) + \eta c_0$, which is a contradiction.\\

{\it Uniqueness} Suppose $u$ is a continuous subsolution  and $w$ a continuous supersolution of \eqref{fb_viscosite} on $S$ satisfying the boundary conditions
$$
u(x,0)\le w(x,0) \quad u(\gamma k,k)\le w(\gamma k,k) \hbox{ for } (x,k) \in S,
$$ 
and the linear growth condition
$$
|u(x,k)|+|w(x,k)|\le C_1 +C_2(x+k) \quad \forall (x,k) \in S,
$$
for some positive constants $C_1$ and $C_2$. We will show by adapting some standard arguments that $u \le w$.
\begin{itemize}
\item[Step 1:] We first construct strict supersolution of \eqref{fb_viscosite} with pertubation of $w$. Set 
$$
h(x,k) = A + B x + Ck + Dxk + Ex^2 + k^2
$$
with

\begin{equation}\label{C1}
A = \frac{1+\mu \bar{\beta} B + \sigma^2 \bar{\beta}^2E  }{r} + C_1
\end{equation}
and
$$
\left\{
\begin{array}{ll}
B &= 2 + \frac{1+C}{\gamma} + \frac{2\mu\bar{\beta}E}{r} \nonumber\\
C &= \frac{\mu\bar{\beta}D}{r}\nonumber\\
D &= 2\gamma E\nonumber\\
E &= \frac{1}{\gamma^2}\nonumber\\
\end{array}
\right.
$$

and define for $\lambda \in [0,1]$ the continuous function on $S$
$$
w^\lambda=(1-\lambda)w+\lambda h.
$$
Because
\[
\left\{
\begin{split}
&\frac{\partial h}{\partial x} - 1 = B+Dk+2Ex - 1 \geq 1 \\
&\gamma \frac{\partial h}{\partial x} - \frac{\partial h}{\partial k} = \gamma (B+Dk+2Ex)-(C+Dx+2k) \geq 1 \\
&\gamma \frac{\partial h}{\partial x} + \frac{\partial h}{\partial k} = \gamma (B+Dk+2Ex)+ (C+Dx+2k)\geq 1 \\
\end{split}
\right.
\]
and
\[
\begin{split}
-\mathcal{L}h &= -(\beta(k)\mu - \alpha((k-x)^+))(B+Dk+2Ex) - \frac{\sigma^2\beta(k)^2}{2}2E + r (A + Bx + Ck + Dxk + Ex^2 + k^2)\\
&\geq (r A - \beta(k)\mu B - \sigma^2 \beta(k)^2E) + (r B - 2 \mu \beta(k)E)x + (rC- \mu \beta(k)D)k\\
&\geq 1.
\end{split}
\]
we have that  
\[
\min\left\{ -\mathcal{L} h, \frac{\partial h}{\partial x} - 1, \gamma \frac{\partial h}{\partial x} - \frac{\partial h}{\partial k}, \gamma \frac{\partial h}{\partial x} + \frac{\partial h}{\partial k}\right\} \geq 1.
\]
which implies that $w^\lambda$ is a strict supersolution of \eqref{fb_viscosite}.
To prove this point, one only needs to take $\bar{x}$ and $\varphi \in C^2$ such that $\bar{x}$ is a minimum of $w^\lambda - \varphi$ and notice that $\bar{x}$ is also a minimum of $w^\lambda - \varphi_2$ with $\varphi_2 = \frac{\varphi - \lambda h}{1-\lambda}$ which allows us to use that $w$ is a viscosity supersolution of \eqref{fb_viscosite}.
\item[Step 2:] In order to prove the strong comparison result, it suffice to show that for every $\lambda \in [0,1]$
$$
\sup_S (u-w^\lambda) \le 0.
$$
Assume by a way of contradiction that there exists $\lambda$ such that
\begin{equation}
\sup_S (u-w^\lambda) > 0.
\label{contradict}
\end{equation}
Because $u$ and $w$ have linear growth, we have
$
\displaystyle{\lim_{||(x,k)||\to+\infty}}(u-w^\lambda)= -\infty.
$\\
Using the boundary conditions
\begin{eqnarray*}
u(x,0)-w^\lambda(x,0) & = & (1- \lambda) (u(x,0) -w(x,0)) + \lambda ( u(x,0) - (A + Bx + E x^2)), \\
 & \le &  \lambda (u(x,0) - (A + Bx + E x^2)),\\
u(\gamma k,k)-w^\lambda(\gamma k,k) & \le & \lambda (u(\gamma k,k)- ( A + (B \gamma + C) k + ( D \gamma + E \gamma^2 +1 )k^2)), 
\end{eqnarray*}
and the linear growth condition, it is always possible to find $C_1$ in Equation \reff{C1} such that  both expressions above are negative
and maximum in Equation \reff{contradict} is reached inside the domain $S$.
\end{itemize}
By continuity of the functions $u$ and $w^\lambda$, there exists a pair $(x_0,k_0)$ with $x_0 \ge \gamma k_0$ such that
$$
M=\sup_S (u-w^\lambda)=(u-w^\lambda)(x_0,k_0).
$$
For $\epsilon > 0$, let us consider the functions
\[
\Phi_\epsilon(x,y,k,l) = u(x,k) - w^\lambda(y,l) - \phi_\epsilon(x,y,k,l)
\]
\[
\phi_\epsilon(x,y,k,l) = \frac{1}{2\epsilon}(|x-y|^2+|k-l|^2) + \frac{1}{4}(|x-x_0|^4+|k-k_0|^4).
\]
By standard arguments in comparison principle of the viscosity solution theory (see Pham \cite{pham:springer} section 4.4.2.), the function $\Phi_\eps$ attains a maximum in $(x_\epsilon,y_\epsilon,k_\epsilon,l_\epsilon)$, which converges (up to a subsequence) to $(x_0, k_0,x_0,k_0)$
when $\eps$ goes to zero. Moreover,
\beq \label{outil1}
\lim_{\epsilon \rightarrow +\infty}\frac{(|x_\epsilon-y_\epsilon|^2+|k_\epsilon-l_\epsilon|^2)}{2\epsilon} \rightarrow 0
\enq
Applying Theorem 3.2 in Crandall Ishii Lions \cite{craishlio92} , we get the existence of symmetric square matrices of size 2 $M_\eps$, $N_\eps$ such that:
\beqs
(p_\eps, M_\eps) \;\in \; J^{2,+} u(x_\eps,k_\eps), \\
(q_\eps, N_\eps) \;\in \; J^{2,-} w^\lambda(y_\eps,l_\eps),
\enqs
and
\beq \label{outil2}
\left ( \begin{array}{cc}
M_\eps & 0 \\
0 & -N_\eps
\end{array} \right)
\; \leq \;  D^2 \phi_\eps(x_\eps, k_\eps,y_\epsilon,l_\epsilon) \; + \; \eps (D^2 \phi_\epsilon(x_\eps, k_\eps,y_\epsilon,l_\epsilon))^2,
\enq
where
$$
p_\eps=D_{x,k}\phi_\epsilon(x_\epsilon,k_\epsilon,y_\epsilon,l_\epsilon) =   \left(\frac{(x_\epsilon - y_\epsilon)}{\epsilon} + (x_\epsilon - x_0)^3,\frac{(k_\epsilon - l_\epsilon)}{\epsilon} + (k_\epsilon - k_0)^3\right),
$$
$$
q_\eps=-D_{y,l}\phi_\epsilon(x_\epsilon,k_\epsilon,y_\epsilon,l_\epsilon) = \left(\frac{(x_\epsilon - y_\epsilon)}{\epsilon},\frac{(k_\epsilon - l_\epsilon)}{\epsilon}\right).
$$
and
\beq
D^2 \phi_\eps(x_\eps, k_\eps,y_\epsilon,l_\epsilon) = \frac{1}{\epsilon}
\left(
\begin{array}{cc}
I_2 & -I_2\\
-I_2 & I_2
\end{array} \right)
+
\left(
\begin{array}{cccc}
3(x_\epsilon - x_0)^2 & 0& 0& 0\\
0 & 3(k_\epsilon - k_0)^2 & 0& 0\\
0&0&0&0\\
0&0&0&0
\end{array} \right)
\enq
so
\[
D^2 \phi_\eps(x_\eps, k_\eps,y_\epsilon,l_\epsilon) \; + \; \eps (D^2 \phi_\epsilon(x_\eps, y_\eps,k_\epsilon,l_\epsilon))^2 = \frac{3}{\epsilon}
\left(
\begin{array}{cc}
I_2 & -I_2\\
-I_2 & I_2
\end{array}
\right)M_\epsilon
\]
\[
+
\left(
\begin{array}{cccc}
9(x_\epsilon - x_0)^2(1 + \epsilon (x_\epsilon - x_0)^2) & 0& 0& 0\\
0 &  9(k_\epsilon - k_0)^2(1 + \epsilon (k_\epsilon - k_0)^2) & 0& 0\\
0&0&0&0\\
0&0&0&0
\end{array} \right)
\]

Equation \eqref{outil2} implies
\beq\label{trace}
\begin{split}
\text{tr} \left(\frac{\sigma^2\beta(k_\epsilon)^2}{2}M_\epsilon - \frac{\sigma^2\beta(l_\epsilon)^2}{2}N_\epsilon\right)\leq &\frac{3\sigma^2}{2\epsilon}(\beta(k_\epsilon)^2 - \beta(l_\epsilon)^2)\\
& + \frac{9\sigma^2 \beta(k_\epsilon)^2}{2} (x_\epsilon - x_0)^2(1+\epsilon(x_\epsilon - x_0)^2)
\end{split}
\enq

Because $u$ and $w^\lambda$ are respectively subsolution and strict supersolution, we have
\begin{equation}\label{ineg_viscosite_u}
\begin{split}
\min\Big[&-\big(\beta (k_\epsilon)\mu - \alpha ((k_\epsilon-x_\epsilon)^+)\big) \Big(\frac{x_\epsilon-y_\epsilon}{\epsilon}+(x_\epsilon-x_0)^3\Big) - \text{tr}(\frac{\sigma^2 \beta(k_\epsilon)^2}{2}M_\epsilon) + r u(x_\epsilon,k_\epsilon),\\
&\frac{x_\epsilon-y_\epsilon}{\epsilon}+(x_\epsilon-x_0)^3 - 1,\\
&\gamma \Big(\frac{x_\epsilon-y_\epsilon}{\epsilon}+(x_\epsilon-x_0)^3\Big) - \Big(\frac{k_\epsilon-l_\epsilon}{\epsilon}+(k_\epsilon-k_0)^3\Big),\\
&\gamma \Big(\frac{x_\epsilon-y_\epsilon}{\epsilon}+(x_\epsilon-x_0)^3\Big) + \Big(\frac{k_\epsilon-l_\epsilon}{\epsilon}+(k_\epsilon-k_0)^3\Big) \Big]
\leq 0
\end{split}
\end{equation}
and
\begin{equation}\label{ineg_viscosite_v}
\begin{split}
\min\Big(&-\big(\beta (l_\epsilon)\mu - \alpha ((l_\epsilon-y_\epsilon)^+)\big) \frac{x_\epsilon-y_\epsilon}{\epsilon} - \text{tr}(\frac{\sigma^2 \beta(l_\epsilon)^2}{2}N_\epsilon) + r w^\lambda(y_\epsilon,l_\epsilon),\\
&\frac{x_\epsilon-y_\epsilon}{\epsilon} - 1,\gamma \frac{x_\epsilon-y_\epsilon}{\epsilon}- \frac{k_\epsilon-l_\epsilon}{\epsilon},\gamma \frac{x_\epsilon-y_\epsilon}{\epsilon}+ \frac{k_\epsilon-l_\epsilon}{\epsilon} \Big)
\geq \lambda .
\end{split}
\end{equation}
\noindent We then distinguish the following four cases:
\begin{itemize}
\item{Case 1.} If $\frac{x_\epsilon-y_\epsilon}{\epsilon}+(x_\epsilon-x_0)^3 - 1 \leq 0$ then we get from \reff{ineg_viscosite_v}, $\lambda + (x_\epsilon - x_0)^3 \leq 0$ yielding a contradiction when $\epsilon$ goes to $0$.
\item{Case 2.} If $\gamma \Big(\frac{x_\epsilon-y_\epsilon}{\epsilon}+(x_\epsilon-x_0)^3\Big) - \Big(\frac{k_\epsilon-l_\epsilon}{\epsilon}+(k_\epsilon-k_0)^3\Big) \leq 0$ then we get from (\ref{ineg_viscosite_v}) $\lambda + \gamma\Big((x_\epsilon-x_0)^3 -(k_\epsilon-k_0)^3\Big) \leq 0$ yielding a contradiction when $\epsilon$ goes to $0$.
\item{Case 3.} If $\gamma \Big(\frac{x_\epsilon-y_\epsilon}{\epsilon}+(x_\epsilon-x_0)^3\Big) + \Big(\frac{k_\epsilon-l_\epsilon}{\epsilon}+(k_\epsilon-k_0)^3\Big) \leq 0$, then we get from (\ref{ineg_viscosite_v}) $\lambda + \gamma \Big((x_\epsilon-x_0)^3 +(k_\epsilon-k_0)^3 \Big)\leq 0$ yielding a contradiction when $\epsilon$ goes to $0$. 
\item{Case 4.} If
\[
-\big(\beta (k_\epsilon)\mu - \alpha ((k_\epsilon-x_\epsilon)^+)\big) \Big(\frac{x_\epsilon-y_\epsilon}{\epsilon}+(x_\epsilon-x_0)^3\Big)   - \text{tr}(\frac{\sigma^2 \beta(k_\epsilon)^2}{2}M_\epsilon) + r u(x_\epsilon,k_\epsilon) \leq 0.
\]
From
\[
-\big(\beta (l_\epsilon)\mu - \alpha ((l_\epsilon-y_\epsilon)^+)\big) \frac{x_\epsilon-y_\epsilon}{\epsilon} - \text{tr}(\frac{\sigma^2 \beta(l_\epsilon)^2}{2}N_\epsilon) + r w^\lambda(y_\epsilon,l_\epsilon) \geq \lambda
\]
we deduce
\[
\begin{split}
\frac{x_\epsilon-y_\epsilon}{\epsilon}\big(\mu(\beta(l_\epsilon)-\beta(k_\epsilon))+\alpha((k_\epsilon-x_\epsilon)^+)-\alpha((l_\epsilon-y_\epsilon)^+)\big)\\
-\text{tr}(\frac{\sigma^2\beta(k_\epsilon)^2}{2}N_\epsilon) + \text{tr}(\frac{\sigma^2\beta(k_\epsilon)^2}{2}N_\epsilon)\\
-\big(\beta (k_\epsilon)\mu - \alpha ((k_\epsilon-x_\epsilon)^+)\big) (x_\epsilon-x_0)^3 \\  
+ r(u(x_\epsilon,k_\epsilon) - w^\lambda(y_\epsilon,l_\epsilon))\leq -\lambda .\\
\end{split}
\]
Using \eqref{trace} we get,
\[
\begin{split}
&\frac{x_\epsilon-y_\epsilon}{\epsilon}\big(\mu(\beta(l_\epsilon)-\beta(k_\epsilon))+\alpha((k_\epsilon-x_\epsilon)^+)-\alpha((l_\epsilon-y_\epsilon)^+)\big)\\
&-\big(\beta (k_\epsilon)\mu - \alpha ((k_\epsilon-x_\epsilon)^+)\big) (x_\epsilon-x_0)^3 + r(u(x_\epsilon,k_\epsilon) - w^\lambda(y_\epsilon,l_\epsilon)) \\
&\leq -\lambda + \frac{3\sigma^2}{2\epsilon}(\beta(k_\epsilon)^2 - \beta(l_\epsilon)^2) + \frac{9\sigma^2 \beta(k_\epsilon)^2}{2} (x_\epsilon - x_0)^2(1+\epsilon(x_\epsilon - x_0)^2).\\  
\end{split}
\]
By sending $\eps$ to zero and using the continuity of $u$,
$w^\gamma _i$, $\alpha$ and $\beta$ we obtain the required contradiction: $r M \leq -\lambda $. 
\end{itemize}
This ends the proof.

\subsection{Proof of the Proposition \ref{Resultgamma=0}}

Because $\beta$ is concave and $\beta'$ goes to $0$ , the existence of $a$ is equivalent to assume 
\begin{equation}
\label{existence-a}
\sigma^2\beta'(0) \ge \frac{\mu}{(1-\delta)}.
\end{equation}
Let us define the function $w_A$ for $A>0$ as the unique solution on $(a,+\infty)$ of the Cauchy problem
$$
\mu\beta(x)w_A'(x) + \frac{\sigma^2\beta(x)^2}{2}w_A''(x) - r w_A(x) = 0
$$
with $w_A(x)=Ax^\delta$ for $0 \le x \le a$ and $w_A$ differentiable at $a$.
\begin{rem}
The Cauchy problem is well defined with the condition $w_A$ differentiable at $a$. Moreover, it is easy to check, using the definition of $a$, that the function $w_A$ is also $C^2$. Because the cost of debt $\alpha$ is high, the shareholders optimally choose not to issue debt but rather adjust costlessly their level of investment.
\end{rem}
\begin{lem}\label{wAcroissante}
For every $A>0$ the function $w_A$ is increasing.
\end{lem}
\begin{pf}
Clearly, $w_A$ is increasing and thus positive on $[0,a]$. Let $c=\min\{x> a\,,w'_A(c)=0\}$. $w_A(c)>0$ because $w_A$ is increasing and positive in a left neighborhood of $c$. Thus, according to the differential equation, we have $w_A''(c)\ge 0$ which implies that $w_A$ is also increasing in a right neighborhood of $c$. Therefore, $w'_A$ cannot become negative.  
\end{pf}
\begin{lem}
\label{concaveGam0}
For every $A>0$, there is some $b_A$ such that $w''_A(b_A)=0$  and $w_A$
is a concave function on $]a, b_A[$.
\end{lem}
\begin{pf}
Assume by a way of contradiction that $w_A''$ does not vanish. 
Using Equations \eqref{delta} and \eqref{equationpoura}, we have 
$$\frac{\sigma^2\beta^2(a)}{2} w_A''(a)=-rAa^\delta.$$
Therefore, we equivalently assume that $w_A''<0$. This implies that $w_A'$ is stricly decreasing and bounded below by $0$ by lemma \ref{wAcroissante} therefore $w_A$ is an increasing concave function. Therefore, $\displaystyle{\lim_{x\to+\infty}}w_A'(x)$ exists and is denoted by $l$. Letting $x \to +\infty$ in the differential equation, we obtain, because $\beta$ has a finite limit,
$$
\frac{\sigma^2\bar{\beta}^2}{2}\lim_{x\to\infty}w_A''(x)=r\lim_{x\to\infty}w_A(x)-\mu\bar{\beta}
l.
$$
Therefore, either $\lim_{x\to\infty}w_A(x)$ is $+\infty$ from which we get a contradiction or finite from which we get $\displaystyle{\lim_{x\to+\infty}}w_A''(x)=0$ by mean value theorem. In the second case, differentiating the differential equation, we have
\begin{eqnarray}
\mu \beta'(x)w_A'(x)+\mu\beta(x)w_A''(x) + \sigma^2\beta'(x)\beta(x)w_A''(x) + \frac{\sigma^2\beta(x)^2}{2}w_A'''(x)-r w_A'(x) = 0
\label{deriveEq}
\end{eqnarray}
Proceeding analogously, we obtain that $\displaystyle{\lim_{x\to+\infty}}w_A'''(x)=0$ and thus $l=0$.
Coming back to the differential equation, we get
$$
0=r\lim_{x\to\infty}w_A(x)
$$
which contradicts that $w_A$ is increasing. Now, define $b_A=\inf\{x \ge a,\,w_A''(x)=0\}$ to conclude.
\end{pf} 
\begin{lem}
There exists $A^*$ such that $w'_{A^*}(b_{A^*})=1$.
\end{lem}
\begin{pf}
For every $A>0$, we have 
\begin{equation}\label{enb_A}
\mu\beta(b_A)w'_A(b_A)=rw_A(b_A).
\end{equation}
Let $A_1=\frac{\mu\bar{\beta}}{ra^\delta}$. Lemma \ref{wAcroissante} yields 
\begin{eqnarray*}
w_{A_1}(b_{A_1})&\ge&w_{A_1}(a)\\
&=&\frac{\mu\bar{\beta}}{r}\\
&\ge&\frac{\mu\beta(b_{A_1})}{r}
\end{eqnarray*}
Therefore, Equation \eqref{enb_A} yields $w'_{A_1}(b_{A_1})\ge 1$.\\
On the other hand, let $A_2=\frac{a^{1-\delta}}{\delta}$. By construction, $w'_{A_2}(a)=1$ and thus $w'_{A_2}(b_{A_2}) \le 1$ by concavity of $w_A$ on $(0,b_A)$. Thus, there is some $A^* \in [\min(A_1,A_2),\max(A_1,A_2)]$ such that $w'_{A^*}=1.$
\end{pf}\\
Hereafter, we denote $b=b_{A^*}$.
\begin{lem}
We have $\mu\beta'(b)\le r$.
\label{boundBeta}
\end{lem}
\begin{pf}
Differentiating the differential equation and plugging $x=b$, we get
$$
\frac{\sigma^2\beta(b)^2}{2}w_A'''(b)+\mu\beta'(b)-r=0
$$
Because $w''_{A^*}$ is increasing in a left neighborhood of $b$, we have $w_A'''(b) \ge 0$ implying the result.

\end{pf}\\
Let us define $$v=\left\{\begin{array}{cc}
w_{A^*}(x)&\quad x\le b\\
x-b+\frac{\mu\beta(b)}{r}&\,x \ge b
\end{array}
\right.
$$
We are in a position to prove the following proposition
\begin{pro}\label{valeurgamma=0highcost}
The shareholders value is $v$.
\end{pro}
\begin{pf}
We have to check that $(v,b)$ satisfies the standard HJB free boundary problem. By construction, $v$ is a $C^2$ concave function on $(0,+\infty)$ satisfying $v'\ge 1$. It remains to check $\max_k\mathcal{L}_kv(x)\le 0$.\\
 For $x >b$, we have
$$
\mathcal{L}_kv(x)=\mu \beta(k) -\alpha((k-x)^+)-\mu \beta(b) -r(x-b).
$$  
If $k\le x$, concavity of $\beta$ and Lemma \ref{boundBeta}   implies
\begin{eqnarray*}
\mathcal{L}_kv(x)&= & \mu(\beta(x)-\beta(b))-r(x-b)\\
&\le & (\mu\beta'(b)-r)(x-b)\\
&\le & 0.
\end{eqnarray*} 
If $k \ge x$, we differentiate $\mathcal{L}_kv(x)$ with respect to $k$ and obtain using again concavity of $\beta$ and convexity of $\alpha$,
$$
\frac{\partial \mathcal{L}_kv(x)}{\partial k}=\mu\beta'(k)-\alpha'(k-x)\le \mu \beta'(0)-\alpha'(0) \le 0.$$ Therefore, $\mathcal{L}_kv(x) \le \mathcal{L}_xv(x)\le 0$.\\
Let $x < b$, because $v$ is concave, the same argument as in the previous lines shows that
$$
\frac{\partial \mathcal{L}_k v(x)}{\partial k}\le 0 \hbox{ for } k \ge x
$$
and therefore 
$$
\max_{k\ge 0}\mathcal{L}_kv(x)=\max_{k \le x}\mathcal{L}_kv(x).
$$
First order condition gives for $0 \leq k < x$
\[
\begin{split}
 \frac{\partial}{\partial k}(\mathcal{L}_kv) &= \mu \beta'(k) v'(x) + \sigma^2\beta'(k)\beta(k)v''(x)\\
&= \beta'(k)[\mu v'(x) + \sigma^2\beta(k)v''(x)].
\end{split}
\]
Thus for $0<x<a$, we have
\[
\frac{\partial}{\partial k}(\mathcal{L}_kv)= \beta'(k)A^*x^{\delta-2}\delta[\mu x+ \sigma^2\beta(k)(\delta-1)]
\]
which gives,
\[
\left\{
\begin{split}
\frac{\partial}{\partial k}(\mathcal{L}0v) > 0\\
 \frac{\partial}{\partial k}(\mathcal{L}_xv) < 0.\\
\end{split}
\right.
\]
Therefore the maximum $k^*(x)$ of $\mathcal{L}_kv(x)$ lies in the interior of the interval $[0,x]$ and satisfies:
\[
\forall 0 <x <a, \beta(k^*(x)) = \frac{\mu x}{\sigma^2(1-\delta)} .
\]
Hence, for $x \le a$, we have by construction
\[
\begin{split}
\max_{0\leq k\leq x}\{\mathcal{L}_kv\} &= \frac{\mu^2x}{\sigma^2(1-\delta)}A^*\delta x^{\delta-1} + \frac{\sigma^2\mu^2x^2}{2\sigma^4(1-\delta)^2}A^*\delta(\delta-1)x^{\delta-2}-r A^*x^\delta\\ 
&=0 .
\end{split}
\]
Now, fix $x \in  (a,b)$. We note that $\displaystyle{\frac{\partial}{\partial k}}(\mathcal{L}_kv)$ has the same sign as $
\mu v'(x) + \sigma^2\beta(k)v''(x)
$ because $\beta$ is strictly increasing. Moreover, because $v$ is concave and $\beta$ increasing, we have
\[
\min_{0 \leq k \leq x} \mu v'(x) + \sigma^2\beta(k)v''(x) = \mu v'(x) + \sigma^2\beta(x)v''(x).
\]
Thus, it suffice to prove $\mu v'(x) + \sigma^2\beta(x)v''(x) \geq 0$ for $x \in (a,b)$ or equivalently because $\beta$ is a positive function that the function $\phi$ defined as
$$\phi(x) = \mu \beta(x) v'(x) + \sigma^2\beta(x)^2v''(x)$$ is positive. We make a proof by contradiction assuming there is some $x$ such that $\phi(x) < 0$. As $\phi(a) = 0$ by Equation \ref{equationpoura} and $\phi(b) > 0$ then there is some $x_1 \in [a,b]$ such that 
\[
\left\{
\begin{split}
\phi(x_1) &< 0\\
\phi'(x_1) &= 0.
\end{split}
\right.
\]
Using the differential equation \eqref{deriveEq} satisfied by $v'$, we obtain
\[
\phi'(x_1) = (2 r - \mu \beta'(x_1))v'(x_1) - \mu \beta(x_1) v''(x_1) = 0
\]
from we deduce
\[
\begin{split}
\phi(x_1) &= \mu \beta(x_1) v'(x_1) + \sigma^2\beta(x_1)^2v''(x_1) \\
&= \mu \beta(x_1) v'(x_1)  + \frac{\sigma^2\beta(x_1)}{\mu} (2 r - \mu \beta'(x_1))v'(x_1) \\
&= \beta(x_1) v'(x_1) ( \mu + \frac{2r\sigma^2}{\mu} - \sigma^2 \beta'(x_1) ).
\end{split}
\]
But $x_1 \geq a$ and thus $\beta'(x_1) \leq \beta'(a)$. Moreover, by  definition of $a$, we have $\sigma^2\beta'(a) \le \frac{ \mu}{(1-\delta)}$. Therefore, Equation \eqref{delta} yields
\[
\begin{split}
\phi(x_1) &\geq \beta(x_1) v'(x_1) \left(\mu + \frac{2r\sigma^2}{\mu} - \frac{\mu}{1-\delta}\right)\\
&\geq \beta(x_1) v'(x_1) \left( \frac{2r\sigma^2}{\mu} - \mu \frac{\delta}{1-\delta}\right)\\
&\geq \beta(x_1) v'(x_1) \left( \frac{2r\sigma^2}{\mu} - \mu \frac{2r\sigma^2}{\mu^2 + 2r\sigma^2}\frac{\mu^2+2r\sigma^2}{\mu^2} \right)\\
&= 0
\end{split}
\]
which is a contradiction.
\end{pf}\\

To complete the characterization of the shareholders value when the cost of debt is high, we have to study the optimal policy when \eqref{existence-a} is not fulfilled. We expect that $a=0$ in that case which means that for all $x$, the manager should invest all the cash in productive assets. Thus we are interested in the solutions to
\begin{equation}\label{equation-E}
\mu\beta(x)w'(x) + \frac{\sigma^2\beta(x)^2}{2}w''(x) - r w(x) = 0
\end{equation}
such that $w(0)=0$.
\begin{pro}\label{dev_lim}
Suppose that the functions $x \rightarrow \frac{x}{\beta(x)}$ and $x \rightarrow \frac{x^2}{\beta(x)^2}$ are analytic in $0$ with a radius of convergence $R$.
The solutions $w$ to Equation \eqref{equation-E} such that $w(0)=0$ are given by
$$
w(x) = \sum_{k=0}^\infty A_k x^{k+y_1}
$$ 
with
\[
\forall k \geq 1, A_k = \frac{1}{-I(k+y_1)}\sum_{j=0}^{k-1}\frac{(j+y_1)p^{(k-j)}(0) + q^{(k-j)}(0)}{(k-j)!}A_j
\]
where the functions $p$ and $q$ are 
\[
\left\{
\begin{split}
p(x) &= \frac{2\mu x}{\sigma^2\beta(x)}\\
q(x) &= - \frac{2r x^2}{\sigma^2 \beta(x)^2}
\end{split}
\right.
\]
the function $I$ is given by
$$
I(y) = \mu\beta'(0)y+\frac{\sigma^2}{2}\beta'(0)^2y(y-1)-r
$$
and $y_1$ is the positive root of $I$
$$
y_1 = \frac{-\mu + \frac{\sigma^2}{2}\beta'(0) + \sqrt{(\mu - \frac{\sigma^2}{2}\beta'(0))^2 + 2r \sigma^2}}{\sigma^2\beta'(0)}.
$$
The radius of convergence of $w$ is at least equal to $R$.
\end{pro}
\begin{pf}
This result is given by the Fuchs' theorem \cite{Na:05}.
\end{pf}\\
Note that the solutions of Equation \eqref{equation-E} vanishing at zero can be written 
$$
w_{A_0}(x)=A_0w_1(x).
$$
If the radius of  convergence of the Frobenius series is finite, then the previously defined function $w_1$  can be extended by use of the Cauchy theorem.

Because $\mu \beta'(0) \ge r$, we have $y_1 <1$. As a consequence, we have
$$
\lim_{x\to 0}w_1'(x)=+\infty \hbox{ and } \lim_{x\to 0}w_1''(x)=-\infty
$$
Thus, proceeding analogously as in Lemma \ref{concaveGam0}, we prove the existence of $b$ such that $w_1''(b)=0$. Because  $w_{A_0}$ is linear in $A_0$, we choose $A_0=A^*=\frac{1}{w'_1(b)}$ to get a concave solution $w^*$ to \eqref{equation-E} with $w^*(0)=0$, $(w^*)'(b)=1$ and $(w^*)''(b)=0$.
We extend $w^*$ linearly on $(b,+\infty)$ as usual to obtain a $C^2$ function on $[0,+\infty[$.
\begin{pro}\label{solution_r1}
The shareholders value is $w^*$.
\end{pro} 
\begin{pf}
It suffices to check that $w^*$ satisfies the free boundary problem. By construction $w^*$ is a $C^2$ concave function on $\mathbb{R}^{+*}$. Because $(w^*)'(b) = 1$, we have
\[
\forall x \in ]0,b], (w^*)'(x) \geq 1
\]
and
\[
\forall x \geq b, (w^*)'(x) = 1.
\]
On $[b,+\infty[$, we have
\[
\begin{split}
\max_{k\geq 0}\{\mathcal{L}_k w^*\} = \max_{k \geq 0}\Big[&\mu\beta(k) - \alpha((k-x)^+) - \mu \beta(b) +  r (b - x)\Big]\\
=\max\Big[&\max_{k\leq x}\mu\beta(k) - \mu \beta(b) +  r (b - x),\\
&\max_{k\geq x}\mu\beta(k) - \alpha(k-x) - \mu \beta(b) +  r (b - x)\Big].\\
\end{split}
\]
Using $\beta$ concave  increasing, $\alpha$ convexe, $\alpha'(0^+)> \mu\beta'(0^+)$, we have
\[
\max_{k\geq 0}\{\mathcal{L}_k w^*\} =\mu\beta(x) - \mu \beta(b) +  r (b - x).
\]
Then using the concavity of $\beta$,
\[
\forall x \geq b,
\max_{k\geq 0}\{\mathcal{L}_kw^*\} \leq 0.
\]
It remains to show that for every $x<b$
\[
\max_{k\geq 0}\{\mathcal{L}_kw^*\} = 0.
\]
Using $\beta$ concave , $\alpha$ convex, $\alpha'(0)> \mu\beta'(0)$ and $w^*$ concave increasing, we have 
\[
\forall k > x, \frac{\partial}{\partial k}(\mathcal{L}_k w^*) = (\mu \beta'(k) - \alpha'(k-x))(w^*)'(x) + \sigma^2\beta'(k)\beta(k)(w^*)''(x) \leq 0.
\]
Thus,
\[
\max_{k\geq 0}\{\mathcal{L}_k(w^*)\} = \max_{0\leq k\leq x}\{\mathcal{L}_k(w^*)\}. 
\]
Moreover,
\[
\begin{split}
\forall 0 < k < x, \frac{\partial}{\partial k}(\mathcal{L}_k(w^*)) &= \mu \beta'(k) (w^*)'(x) + \sigma^2\beta'(k)\beta(k)(w^*)''(x)\\
&= \beta'(k)[\mu (w^*)'(x) + \sigma^2\beta(k)(w^*)''(x)].
\end{split}
\]
We expect
\[
\forall x \in ]0,b], \forall k \leq x, \frac{\partial}{\partial k}(\mathcal{L}_k(w^*)) \geq 0.
\]
Notice that $\beta'(k) \geq 0$ and
\[
\min_{0 \leq k \leq x} \mu (w^*)'(x) + \sigma^2\beta(k)(w^*)''(x) = \mu (w^*)'(x) + \sigma^2\beta(x)(w^*)''(x)
\]
because $(w^*)''(x) \leq 0$ and $\beta$ is increasing.
Thus it is enough to prove for every $x<b$,
\[
 \mu (w^*)'(x) + \sigma^2\beta(x)(w^*)''(x) \geq 0
\]
or equivalently, using $\beta \geq 0$,
 
\[
\phi(x) = \mu \beta(x) (w^*)'(x) + \sigma^2\beta(x)^2 (w^*)''(x) \geq 0
\]

for $x<b$.
We make a proof by contradiction assuming the existence of $x$ such that $\phi(x) < 0$. In a neighborhood of $0$, we have  
\[
(w^*)'(x) \sim A^* y_1x^{y_1 -1} 
\]
and
\[
(w^*)''(x) \sim A^* y_1(y_1-1)x^{y_1-2}
\]
From which we deduce because $\beta(x)x^{y_1-1}\le \beta'(0)x^{y_1}$,
\[
\lim_{x\rightarrow 0} \beta(x)(w^*)'(x) = 0
\]
\[
\lim_{x\rightarrow 0} \beta(x)^2(w^*)''(x) = 0
\]
yielding
\[
\lim_{x\rightarrow 0} \phi(x) = 0.
\]
But $\phi(b) > 0$ thus there is $x_1 \in ]0,b[$ such that 
\[
\left\{
\begin{split}
\phi(x_1) &< 0\\
\phi'(x_1) &= 0.
\end{split}
\right.
\]
Using the derivative of Equation \eqref{equation-E} 
\[
\phi'(x_1) = (2 r - \mu \beta'(x_1))(w^*)'(x_1) - \mu \beta(x_1) (w^*)''(x_1) = 0
\]
from which we deduce :
\[
\begin{split}
\phi(x_1) &= \mu \beta(x_1) (w^*)'(x_1) + \sigma^2\beta(x_1)^2(w^*)''(x_1) \\
&= \mu \beta(x_1) (w^*)'(x_1)  + \frac{\sigma^2\beta(x_1)}{\mu} (2 r - \mu \beta'(x_1))(w^*)'(x_1) \\
&= \beta(x_1) (w^*)'(x_1) ( \mu + \frac{2r\sigma^2}{\mu} - \sigma^2 \beta'(x_1) ).
\end{split}
\]
Now, remember that $x_1 > 0$ and thus using the concavity of $\beta$, we have $\beta'(x_1) \leq \beta'(0)$. Furthermore, $\beta'(0) \leq \frac{\mu^2+2r \sigma^2}{\sigma^2\mu}$ when Equation \eqref{equationpoura} is not fulfilled. Hence,
\[
\begin{split}
\phi(x_1) &\geq \beta(x_1) (w^*)'(x_1) \left(\mu + \frac{2r\sigma^2}{\mu} - \frac{\mu^2 + 2r \sigma^2}{\mu}\right)\\
&\geq 0
\end{split}
\]
which yields to a contradiction and ends the proof. 
\end{pf}

\end{document}